\documentclass[12pt,a4paper]{article}
\usepackage[utf8]{inputenc}
\usepackage[lmargin=1in, rmargin=1in, tmargin=1in, bmargin=1in]{geometry}
\usepackage{setspace}
\usepackage{titling}
\usepackage{lipsum}
\usepackage{comment}
\usepackage{physics}

\usepackage[
    protrusion=true,
    activate={true,nocompatibility},
    final,
    tracking=true,
    kerning=true,
    spacing=true,
    factor=1100]{microtype}
\microtypecontext{spacing=nonfrench}

\usepackage{longtable} 

\usepackage{tikz}
\usetikzlibrary{shapes}
\usetikzlibrary{matrix}
\usepackage[normalsize]{subfigure}
\usetikzlibrary{backgrounds,automata}
\usepackage[
singlelinecheck=false 
]{caption}
\usepackage{float}

\usepackage{amsmath,amsfonts,amssymb,bbm}
\usepackage{mathrsfs}

\usepackage{mathtools, nccmath}
\usepackage[english]{babel}
\usepackage{blindtext}

\usepackage{amsthm} 

\newtheoremstyle{normalfontstyle} 
  {}                              
  {}                              
  {\normalfont}                   
  {}                              
  {\bfseries}                     
  {.}                             
  { }                             
  {}                              

\theoremstyle{normalfontstyle}

\newtheorem{lemm}{Lemma}
\newtheorem{prop}{Proposition}

\newtheorem{assu}{Assumption}

\usepackage{natbib}

\usepackage{graphicx}
\newcommand{\indep}{\rotatebox[origin=c]{90}{$\models$}}  

\newcommand{\Cov}{\operatorname{Cov}}
\newcommand{\Var}{\operatorname{Var}}
\newcommand{\E}{\operatorname{E}}

\def\one{\mathbbm{1}}

\DeclareMathOperator{\Pro}{Pr}
\def\one{\mathbbm{1}}

\newcommand{\defeq}{\vcentcolon=}

\usepackage{changepage}  

\usepackage[colorlinks=true, allcolors=blue, hyperfootnotes=false]{hyperref}  

\usepackage{makecell}

\allowdisplaybreaks

\usepackage{etoolbox}
\makeatletter

\pretocmd{\NAT@citex}{%
  \let\NAT@hyper@\NAT@hyper@citex
  \def\NAT@postnote{#2}%
  \setcounter{NAT@total@cites}{0}%
  \setcounter{NAT@count@cites}{0}%
  \forcsvlist{\stepcounter{NAT@total@cites}\@gobble}{#3}}{}{}
\newcounter{NAT@total@cites}
\newcounter{NAT@count@cites}
\def\NAT@postnote{}

\def\NAT@hyper@citex#1{%
  \stepcounter{NAT@count@cites}%
  \hyper@natlinkstart{\@citeb\@extra@b@citeb}#1%
  \ifnumequal{\value{NAT@count@cites}}{\value{NAT@total@cites}}
    {\ifNAT@swa\else\if*\NAT@postnote*\else%
     \NAT@cmt\NAT@postnote\global\def\NAT@postnote{}\fi\fi}{}%
  \ifNAT@swa\else\if\relax\NAT@date\relax
  \else\NAT@@close\global\let\NAT@nm\@empty\fi\fi
  \hyper@natlinkend}
\renewcommand\hyper@natlinkbreak[2]{#1}

\patchcmd{\NAT@citex}
  {\ifNAT@swa\else\if*#2*\else\NAT@cmt#2\fi
   \if\relax\NAT@date\relax\else\NAT@@close\fi\fi}{}{}{}
\patchcmd{\NAT@citex}
  {\if\relax\NAT@date\relax\NAT@def@citea\else\NAT@def@citea@close\fi}
  {\if\relax\NAT@date\relax\NAT@def@citea\else\NAT@def@citea@space\fi}{}{}

\makeatother

\providecommand{\keywords}[1]
{
  \small	
  \textbf{\textit{Keywords---}} #1
}

\title{\large Detecting and Understanding the Difference between Natural Mediation Effects and Their Randomized Interventional Analogues}
\author{Ang Yu\thanks{Department of Sociology, University of Wisconsin-Madison. Email: ayu33@wisc.edu}, Li Ge\thanks{Biostatistics and Research Decision Sciences, Merck \& Co., Inc., Rahway, NJ, and Department of Biostatistics and Medical Informatics, University of Wisconsin-Madison}, and Felix Elwert\thanks{Department of Sociology, Department of Biostatistics and Medical Informatics, and Department of Population Health Sciences, University of Wisconsin-Madison}}
\date{Apr 17, 2025}

\begin{document}

\maketitle


\begin{abstract}
In causal mediation analysis, the natural direct and indirect effects (natural effects) are nonparametrically unidentifiable in the presence of treatment-induced confounding, which motivated the development of randomized interventional analogues (RIAs) of the natural effects. Being easier to identify, the RIAs are becoming widely used in practice. However, applied researchers often interpret RIA estimates as if they were the natural effects, even though the RIAs can be poor proxies for the natural effects. This calls for practical and theoretical guidance on when the RIAs differ from or coincide with the natural effects. We develop the first empirical test to detect the divergence between the natural effects and their RIAs under the weak assumptions sufficient for identifying the RIAs and illustrate the test using the Moving to Opportunity Study. We also provide new theoretical insights on the relationship between the natural effects and the RIAs both using a covariance formulation and from a structural equation perspective. This analysis also reveals previously undocumented connections between the natural effects, the RIAs, and estimands in instrumental variable analysis and Wilcoxon-Mann-Whitney tests. 
\end{abstract}  


\section{Introduction}
\subsection{Background}
Causal mediation analysis explains the mechanisms of a total causal effect by decomposing it into direct and indirect effects in terms of one or more mediators. The direct effect is the component of the total effect that does not operate through the mediators of interest, and the indirect effect is the component that does \citep{vanderweele_explanation_2015, hong2015causality, nguyen_clarifying_2022}. As a central task in the social and health sciences, causal mediation analysis is widely used in applied research. 

We adopt the conventional notation of causal mediation analysis. 
$Y$ is the observed outcome, $A$ is a binary treatment (or any pair of two values for a multivalued treatment) labeled $\{0,1\}$, and $M$ is a vector of mediators. $Y_a$ and $M_a$ are, respectively, the potential values of $Y$ and $M$ under the assignment of treatment value $a$. We further define two sets of confounders that may be empty, $C$ is a set of pre-treatment confounders, and $L$ is a set of post-treatment confounders. Unless otherwise stated, we allow all $L$ to be treatment-induced confounders, i.e., confounders of the $M$-$Y$ relationship that are affected by the treatment. Figure \ref{fig:dag} illustrates the relationship between variables, when any variable may affect any temporally subsequent variables.

\begin{figure}[ht]
    \centering
\begin{tikzpicture}[scale = 1] 
    \node at (0,0) {};
    \node at (1,1) {};
    
    \node[anchor = center, align = center] (a) at (0,-1) {$A$};
    \node[anchor = center, align=center] (m) at (3,-1) {$M$}; 
    \node[anchor = center, align=center] (y) at (6,-1) {$Y$};   
    \node[anchor = center, align=center] (c) at (-3,-1) {$C$};
    \node[anchor = center, align=center] (l) at (3, -3) {$L$};
    
    \draw[->, thick] (a) to (m);
    \draw[->, thick] (m) to (y);
    \draw[->, thick] (a) to [bend left = 20] (y);  
    
    \draw[->, thick] (c) to  (a);  
    \draw[->, thick] (c) to [bend right = 12] (l);  
    \draw[->, thick] (c) to [bend left = 30] (m);  
    \draw[->, thick] (c) to [bend left = 30] (y);  
    \draw[->, thick] (a) to (l); 
    \draw[->, thick] (l) to (m); 
    \draw[->, thick] (l) to (y); 
\end{tikzpicture}
\caption{Variable set-up in causal mediation analysis}
    \label{fig:dag}
\end{figure}

The canonical approach of causal mediation analysis decomposes the total effect (TE) into the natural indirect effect (NIE) and the natural direct effect (NDE) \citep{robins1992,pearl_direct_2001}. 

\begin{equation*}
    \underbrace{\E(Y_1-Y_0)}_{\text{TE}} = \underbrace{\E(Y_{1,M_1}-Y_{0,M_0})}_{\text{TE}}= \underbrace{\E(Y_{1,M_1}-Y_{1,M_0})}_{\text{NIE}}  + \underbrace{\E(Y_{1,M_0}-Y_{0,M_0})}_{\text{NDE}},
\end{equation*}
where $Y_{a,M_{a'}}$ denotes the potential outcome of $Y$ under the assignment of treatment $a$ and the mediator value that would be realized under the assignment of treatment $a'$. 
The NIE is defined by fixing treatment assignment at 1 and varying the mediator assignment from $M_0$ to $M_1$, capturing the part of the total effect that operates through $M$. The NDE is defined by varying the treatment assignment from 0 to 1 but holding mediator assignment at the baseline mediator value, capturing the part of the total effect that does not operate through $M$.
Importantly, the natural effects (NIE and NDE) aggregate individual-level causal mechanisms, as they are based on individual-level potential mediators, $M_1$ and $M_0$.

The natural effects are notoriously difficult to identify. Without parametric assumptions, they are unidentifiable when there exists any treatment-induced confounder $L$, regardless of whether $L$ is observed \citep{green_semantics_2003,avin_identiability_2005}. 
Therefore, the application of natural effects is challenging in many empirical settings, as ruling out $L$ altogether is often impossible, and parametric assumptions are often hard to justify.

Motivated by the difficulty of identifying the natural effects, statisticians have proposed an alternative decomposition whose nonparametric identification does not require the absence of treatment-induced confounders \citep{vanderweele_effect_2014}. This alternative decomposition is based on the randomized interventional analogues (RIA) of the TE, the NIE, and the NDE, namely the TE$^R$, the NIE$^R$, and the NDE$^R$:
\begin{equation*}
    \underbrace{\E(Y_{1,G_1}-Y_{0,G_0})}_{\text{TE}^R}= \underbrace{\E(Y_{1,G_1}-Y_{1,G_0})}_{\text{NIE}^R}  + \underbrace{\E(Y_{1,G_0}-Y_{0,G_0})}_{\text{NDE}^R},
\end{equation*}
where $G_{a'}$ is a value randomly drawn from the mediator distribution that would be realized under the assignment of treatment value $a'$ given $C$, and $Y_{a, G_{a'}}$ is the potential outcome of $Y$ under the assignment of the treatment value $a$ and the mediator value $G_{a'}$. Clearly, the RIAs differ from the natural effects in mediator assignments: instead of $M_1$ and $M_0$, the mediator assignments for the RIAs are $G_1$ and $G_0$. 

Seen as much less demanding and more widely applicable than the natural effects, the RIAs have become popular in empirical research.
In practice, applied researchers frequently estimate the RIAs as proxies of the natural effects. In fact, the RIA estimates are often interpreted as if they were estimates of the natural effects. \citet{sarvet_interpretational_2023} reviewed 16 applied studies that estimate RIAs, all of which contain interpretive statements that elide the difference between the RIAs and the natural effects. The methodological literature has encouraged this ambiguity. First, the RIAs are named as analogues to begin with \citep{vanderweele_effect_2014}. Second, \citet{vanderweele_mediation_2017} write that “it will only be in extremely unusual settings that the interventional analogue is non-zero, with there being no natural indirect effects.” 


However, there are reasons to suspect that the RIAs can be poor proxies of the natural effects. Unlike the natural effects, they are not individual-level explanatory mechanisms. Formalizing this intuition, \citet{miles_causal_2023} proposes a set of null criteria that valid indirect effect measures should satisfy and shows that the NIE is valid by these criteria while the NIE$^R$ is not. In particular, the NIE$^R$ can be nonzero even if the mediator does not ``mediate'' the treatment effect for any individual (more detail in Section \ref{sec: cov_binary}). In addition, it has been frequently noted in the methodological literature that the NIE$^R$ and the NDE$^R$ do not generally sum to the TE, which is problematic because the canonical task of causal mediation analysis is to understand the TE \citep{vansteelandt_interventional_2017,nguyen_clarifying_2021}. 

\begin{figure}
    \centering
    \includegraphics[width=\textwidth]{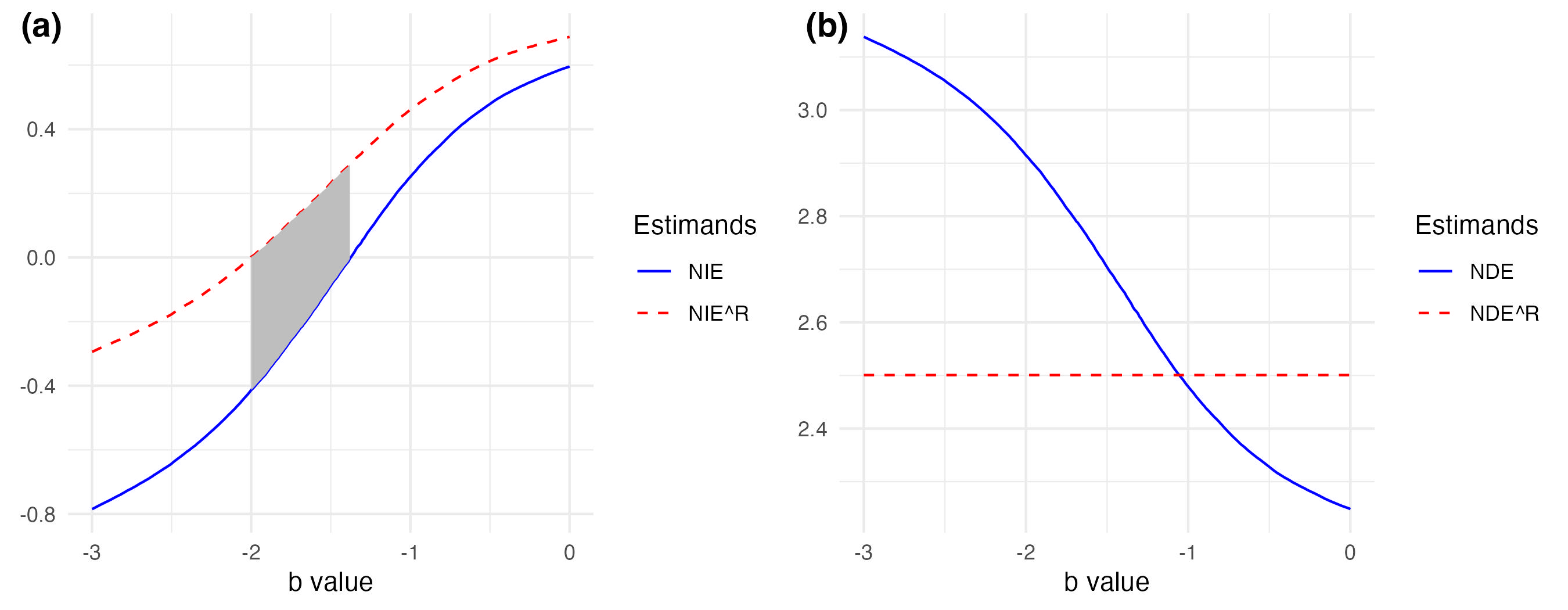}
    \caption{Illustration of possible divergence between the natural mediation estimands and their RIAs. Panel (a) depicts NIE and NIE$^R$, and Panel (b) depicts NDE and NDE$^R$. The DGP is as follows. $L \sim \mathcal{N}(A, 1)$; $M \sim \text{expit}(A+L+bAL)$; $Y \sim \mathcal{N}(A+L+M+LM,1)$. The $x$ axis is the coefficient of the $A$-$L$ interaction in the generative model for $M$. In Panel (a), the shaded area indicates sign reversal between NIE and NIE$^R$.}
    \label{fig:sim}
\end{figure}

In contrast to the violation of null criteria, which focuses on a knife-edge scenario, we draw attention to possible quantitative differences between the natural effects and the RIAs in a wide range of data generating processes (DGPs). These quantitative differences may be large and even involve sign reversal. In the illustration of Figure \ref{fig:sim}, data are simulated according to a set of very simple and seemingly innocuous DGPs. By varying one parameter of the DGP, we observe areas of substantial divergence and even sign reversal, where the RIAs can hardly be used to draw conclusions about the natural effects.   

Therefore, it is natural to ask when the natural effects differ from their RIAs. If they are identical or at least close to each other, then it might be warranted to interpret estimates of the RIAs as the natural effects, as is common in empirical research. 
Conversely, if they substantially differ, then more caution and precision in interpretation is called for. In this paper, we answer this question by developing the first empirical test for the difference between the natural effects and their RIAs and by introducing two complementary theoretical perspectives to explain when and how they differ.

\subsection{Contributions}
We propose a novel test for the differences between the NIE, the NDE, and their respective RIAs. The empirical testability of these differences may be surprising, because under the standard assumptions for identifying the NIE and the NDE, the natural effects necessarily coincide with their RIAs \citep[][p.921]{vanderweele_mediation_2017}. On the other hand, under the standard assumptions identifying the NIE$^R$ and the NDE$^R$, the NIE and the NDE are unidentified. Thus, it may appear that under no set of common assumptions can one test the differences. However, our test is made possible by leveraging two simple facts. First, the TE and the TE$^R$ are identified under the standard assumptions for the NIE$^R$ and the NDE$^R$. Second, when $\text{TE} -\text{TE}^{R} \neq 0$, it is necessarily the case that either $\text{NIE} \neq \text{NIE}^{R}$ or $\text{NDE} \neq \text{NDE}^{R}$. Hence, instead of hoping that ``the natural and interventional effects may coincide empirically'' \citep[][p.2]{Loh_2020}, analysts can test their divergence by testing $\text{TE} -\text{TE}^{R}=0$ under the weak identifying assumptions that are sufficient for the RIAs but not the natural effects. 

We also theoretically clarify and illustrate the substantive conditions under which the natural effects differ from or coincide with their RIAs. We do so from a nonparametric covariance perspective and a structural equations perspective. First, we derive a covariance-based representation of the differences between the natural effects and their RIAs. Second, we derive parametric constraints on the structural equations generating the data under which the the natural effects will coincide with the RIAs. These two novel perspectives provide exact and intuitive insights on the substantive mechanisms underpinning the relationship between the natural effects and the RIAs. We thus go beyond and demystify Miles' (\citeyear{miles_causal_2023}) results, which prove the null criteria violation of the NIE$^R$ using a specific numerical counterexample. Additionally, we also present the relationship between the natural effects and the organic effects of \citet{lok_defining_2016} from the covariance perspective. 

The remainder of this paper is organized as follows. In Section 2, we review the standard assumptions in causal mediation analysis. In Section 3, we present our empirical test for the differences between the natural effects and the RIAs and apply it to the Moving to Opportunity (MTO) study. Sections 4 and 5, respectively, introduce the covariance perspective and the structural equation perspective. Section 6 discusses related estimands, including those in the instrumental variable (IV) settings and those underlying the Wilcoxon-Mann-Whitney tests. We present novel results that unify causal mediation analysis with these other fields of causal inference. Section 7 concludes. 
All proofs are collected in the appendix. R code for simulating Figure \ref{fig:sim} and empirical data analysis in Section \ref{sec: test} can be found at \href{https://github.com/ang-yu/diff_naturals_rias}{https://github.com/ang-yu/diff\_naturals\_rias}.

\section{Review of conventional mediation assumptions}

The literature on causal mediation analysis predominately relies on combinations of the following five assumptions.

\begin{assu}[Consistency]
    $f(M_a \mid C,a)=f(M \mid C,a)$ and $\E(Y_{a,m} \mid C,a,L,m)=\E(Y \mid C,a,L,m)$, for all $a$ and $m$, where $f(\cdot)$ is the density function. \label{assu: consistency}
\end{assu}
\begin{assu}[Ignorability of $A$ conditional on $C$]
    $Y_{a,m} \indep A \mid C$ for all $a$ and $m$; $M_{a} \indep A \mid C$ for all $a$. \label{assu: A}
\end{assu}  
\begin{assu}[Ignorability of $M$ conditional on $C,A,L$]
    $Y_{a,m} \indep M \mid C, A=a, L$ for all $a$ and $m$. \label{assu: MY_CL}
\end{assu}
\begin{assu}[Ignorability of $M$ conditional on $C,A$]
    $Y_{a,m} \indep M \mid C, A=a$ for all $a$ and $m$. \label{assu: MY_C}
\end{assu}
\begin{assu}[Cross-world Independence]
    $Y_{a,m} \indep M_{a'} \mid C$ for all $a$, $a'$, and $m$. \label{assu: cross}
\end{assu}
Assumption \ref{assu: consistency} links the potential values of the mediator $M$ and the outcome $Y$ to their observed values. Assumption \ref{assu: A} requires the treatment $A$ to be ignorable conditional on baseline confounders $C$. Assumption \ref{assu: MY_CL} states that $M$ is conditionally ignorable given both $C$ and post-treatment confounders $L$, as well as the treatment. Assumption \ref{assu: MY_C} imposes conditional ignorability of the mediator given only baseline confounders and the treatment, which is stronger than Assumption \ref{assu: MY_CL}.
Finally, Assumption \ref{assu: cross} requires the conditional independence between the potential outcomes $Y_{a,m}$ and potential mediators $M_{a'}$ under two possibly different treatment assignments $a$ and $a'$, hence its name (cross-world independence). 

In the literature, Assumptions \ref{assu: consistency}, \ref{assu: A}, and \ref{assu: MY_CL}, are the standard identifying assumptions for the RIAs \citep{vanderweele_effect_2014}, while Assumptions \ref{assu: consistency}, \ref{assu: A}, \ref{assu: MY_C}, and \ref{assu: cross} are the standard assumptions for identifying the NIE and the NDE (Pearl, \citeyear{pearl_direct_2001}; VanderWeele, \citeyear{vanderweele_explanation_2015}, p.463-4; See Imai, Keele, and Yamamoto, \citeyear{imai_identification_2010} for a slightly stronger version). Notably, the cross-world independence assumption requires the absence of any post-treatment confounder of the mediator-outcome relationship ($L=\varnothing$) \citep{green_semantics_2003,avin_identiability_2005,andrews_insights_2021}. Hence, it is clear that the standard assumptions for the RIAs are weaker, as they allow for the existence of post-treatment confounders, $L$. Finally, when the cross-world independence assumption holds, the natural effects are necessarily equivalent to their RIAs \citep{vanderweele_mediation_2017}.

\section{Empirical test}
\label{sec: test}
We propose to use the empirical estimate of $\text{TE} -\text{TE}^{R}$ as a test statistic for the divergence between the NIE and the NIE$^R$ and between the NDE and the NDE$^R$. This test relies on the fact that if $\text{TE} -\text{TE}^{R} \neq 0$, it is necessarily true that either $\text{NIE} \neq \text{NIE}^{R}$ or $\text{NDE} \neq \text{NDE}^{R}$, or both. Thus, if we reject the null hypothesis that $\text{TE} - \text{TE}^{R} = 0$, we also reject the composite null hypothesis that $\text{NIE} = \text{NIE}^{R}$ and $\text{NDE} = \text{NDE}^{R}$. Therefore, our test is a falsification test for the composite null hypothesis. The composite null hypothesis is practically relevant, as empirical work always presents both $\text{NIE}^{R}$ and $\text{NDE}^{R}$, requiring both to be correctly interpreted. 
In addition, since $\vert \text{TE} - \text{TE}^{R} \vert \leq \vert \text{NIE} - \text{NIE}^{R} \vert + \vert \text{NDE} - \text{NDE}^{R} \vert$ by the triangle inequality, $\vert \text{TE} - \text{TE}^{R} \vert$ also provides a lower bound for the sum of the absolute differences between the NIE and the NIE$^{R}$ and between the NDE and the NDE$^{R}$.

\begin{figure}
    \centering
    \includegraphics[width=0.6\textwidth]{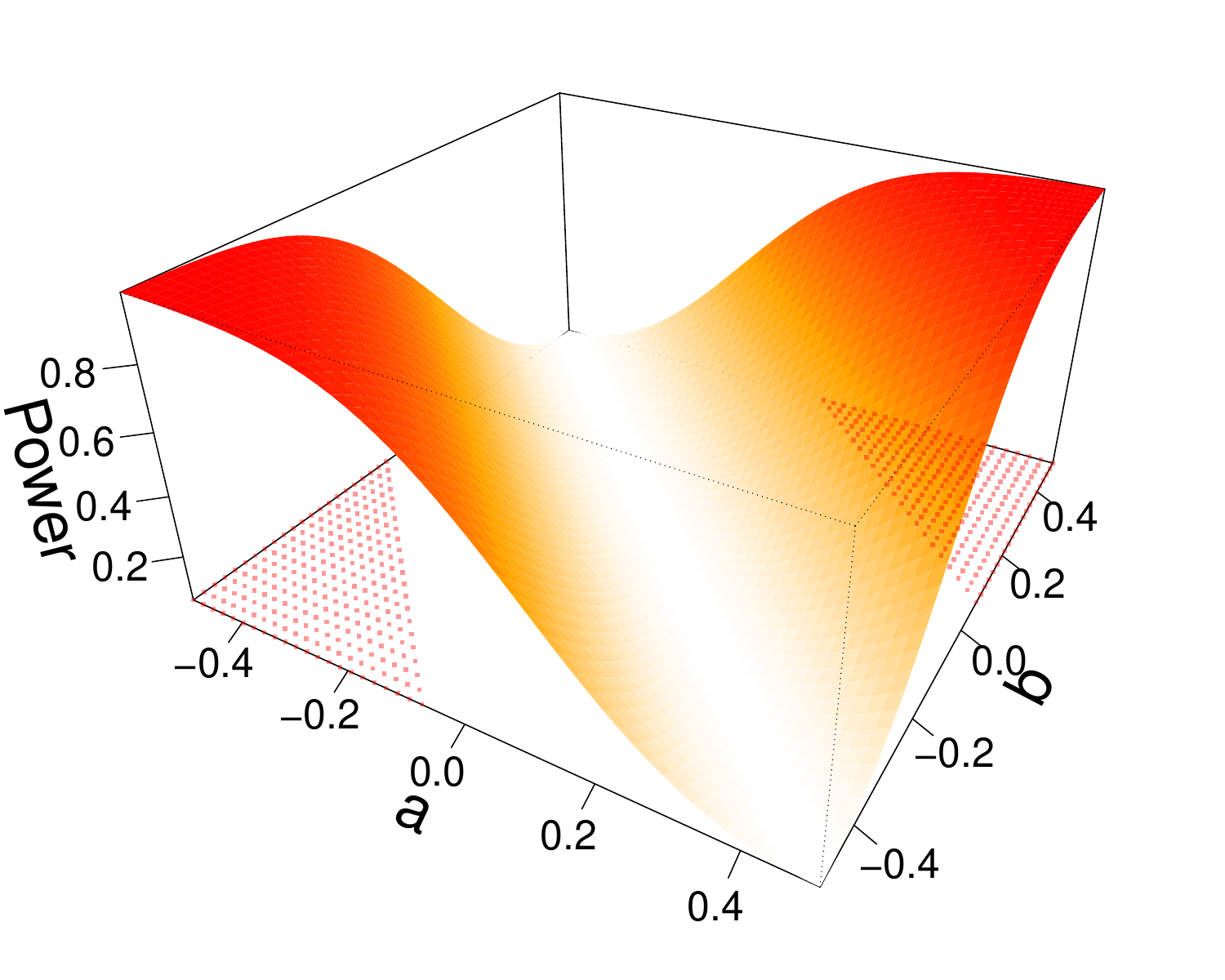}
    \caption{Illustration of test power as a function of the differences between the natural effects and the RIAs under the alternative hypothesis. We respectively vary $a=\text{NIE}-\text{NIE}^{R}$ and $b=\text{NDE}-\text{NDE}^{R}$ in the alternative hypothesis from -0.5 to 0.5. We fix other parameters of power formula (\ref{equ:power}) as such: $\alpha=0.05$, $n=25$, and $\sigma=1$. The areas where the power exceeds 0.8 are  highlighted on the x-y plane.}
    \label{fig:power}
\end{figure}

As a limitation of our test, note that $\text{TE} -\text{TE}^{R} = 0$ does not imply $\text{NIE} = \text{NIE}^{R}$ and $\text{NDE} = \text{NDE}^{R}$, as $\text{NIE} - \text{NIE}^{R}$ and $\text{NDE}-\text{NDE}^{R}$ may be both nonzero but exactly cancel each other out. More broadly, the power of our test depends on the extent to which they cancel out. Consider the following null and alternative hypotheses:
\begin{align*}
    H_0&: \text{NIE} - \text{NIE}^{R}=0 \text{ and } \text{NDE} - \text{NDE}^{R}=0 \\
    H_1&: \text{NIE}-\text{NIE}^{R}=a \text{ and } \text{NDE} - \text{NDE}^{R}=b.
\end{align*}
Assuming that our estimate of $\text{TE} -\text{TE}^{R}$ is (asymptotically) normally distributed, then the power of the two-sided test with a level of $\alpha$ is

\begin{equation}
    \Phi\left( -z_{1 - \alpha/2} + \frac{\sqrt{n}(a+b)}{\sigma} \right) + 1 - \Phi\left( z_{1 - \alpha/2} + \frac{\sqrt{n}(a+b)}{\sigma} \right), \label{equ:power}
\end{equation}
where $\Phi(\cdot)$ is the CDF of the standard normal distribution, $z_{1 - \alpha/2}$ is the $1-\alpha/2$ quantile of the standard normal distribution, and $\sigma$ is the standard deviation of the estimate of $\text{TE} -\text{TE}^{R}$. As Figure \ref{fig:power} shows, the more $a$ and $b$ offset each other, in the sense that $a \rightarrow -b$, the lower the power. Conversely, the power increases as $\vert a+b \vert$ becomes larger.

Under Assumptions \ref{assu: consistency}, \ref{assu: A} and \ref{assu: MY_CL}, $\text{TE} -\text{TE}^{R}=\E(Y_{1})-\E(Y_{0})-\E(Y_{1,G_1}) + \E(Y_{0,G_0})$ is identified by the functionals below \citep{vanderweele_effect_2014}.

\begin{align*}
\E(Y_{a}) &= \iint y f(y \mid c,a) f(c) \dd y \dd c \\
    \E(Y_{a,G_a}) &= \iiiint y f(y \mid c,a,l,m) f(m \mid c,a) f(l \mid c,a) f(c)  \dd y \dd m \dd l \dd c.
\end{align*}
Hence, importantly, our test parameter, $\text{TE} -\text{TE}^{R}$, is nonparametrically identifiable even when there are treatment-induced confounders, $L$, and Assumption \ref{assu: cross} is invalid. This is because although the NIE and the NDE are not nonparametrically identifiable under treatment-induced confounding, their sum is. 

The task now is to estimate $\text{TE} -\text{TE}^{R}$. This can be done using various estimators of TE and TE$^R$. For TE, various regression, weighting, or efficient influence function (EIF)-based estimators are well-known and can be found in standard textbooks of causal inference \citep[e.g.,][]{imbens2015causal,hernan_causal_2020}. For $\text{TE}^{R}$, \citet{vanderweele_effect_2014} and \citet{wodtke_effect_2020} introduced parametric estimators via weighting and regression, respectively. These estimators are prone to misspecification biases, because they require the functional form assumptions of all component models be satisfied. In response to the disadvantage of parametric estimators, \citet{diaz_nonparametric_2021} and \citet{rudolph_practical_2024} developed nonparametric estimators based on the EIF of $\text{TE}^{R}$. These estimators do not impose  functional form assumptions and are robust to inconsistent estimation of some component models. However, these estimators are only computationally tractable when either $L$ or $M$ is discrete and low-dimensional. 

We recommend a Riesz Regression (RR) approach built on the recent works of \citet{chernozhukov_2024} and \citet{liu_general_2024}, who developed estimators for TE and TE$^R$ that can be readily repurposed for $\text{TE} -\text{TE}^{R}$. The RR approach has multiple desirable properties. As an EIF-based approach, it is nonparametric and doubly robust, hence not prone to misspecification in functional form. This approach also attains semiparametric efficiency and asymptotic normality under relatively mild conditions. Furthermore, compared with previous EIF-based estimators of $\text{TE}^{R}$ \citep{diaz_nonparametric_2021, rudolph_practical_2024}, the RR approach can very generally accommodate arbitrary numbers and types of $L$ and $M$ variables. For the technical details of the RR approach, we refer readers to \citet{chernozhukov_2024} and \citet{liu_general_2024}. Practically, we extend the \texttt{\{crumble\}} R package developed by \citet{Williams_2024} to facilitate easy implementation of our test. The extended R package is available for download from the Github repository \href{https://github.com/ang-yu/ria_test}{https://github.com/ang-yu/ria\_test}.

\subsection{Empirical illustration}
We apply our test to a mediation analysis of the Moving to Opportunity (MTO) study, a large-scale longitudinal randomized control trial conducted by the Department of Housing and Urban Development of the United States \citep{ludwig_neighborhood_2013, kling_neighborhood_2005}. We follow the conceptual set-up of \citet{rudolph_helped_2021} and \citet{rudolph_practical_2024}, who estimated the RIAs.\footnote{Due to lack of access to the restricted-use dataset, we follow their variable and sample choices only approximately. Hence, our estimates should be regarded as purely illustrative.} The treatment ($A$) is a binary indicator of whether or not a family living in a high-poverty neighborhood was randomized to receive a Section 8 housing voucher that allowed them to move to a less poor neighborhood. We consider two mediators ($M$) measured between 10 and 15 years of follow up, neighborhood poverty and the number of residential moves. The outcome ($Y$) is a composite score of mental health \citep{ludwig_neighborhood_2013}. For causal identification, we account for a post-treatment confounder ($L$) which is whether the family used the voucher to move within the 90 days allotted. We also account for 12 baseline confounders ($C$), which capture baseline household socioeconomic and demographic characteristics, as well as neighborhood-related perceptions and aspirations. 

\begin{table}[]
    \centering
    \begin{tabular}{c c c}
        Estimand & Estimate & 95\% Confidence Interval \\
        \hline
        TE & 0.0495 & (0.0386, 0.0604) \\
        TE$^R$ & 0.0619 & (0.0511, 0.0726) \\
        $\text{TE}-\text{TE}^R$ & -0.0123 & (-0.0173, -0.0074) \\
        NIE$^R$ & 0.0287 & (0.0194, 0.0380) \\
        NDE$^R$ & 0.0332 & (0.0296, 0.0367) \\
        \hline
    \end{tabular}
    \caption{Empirical estimates from the MTO study. $N=3270$. The treatment is the receipt of a randomized housing voucher. The mediators are neighborhood poverty and the number of residential moves. The outcome is mental health, which is standardized to have unit variance. Estimation is done by the Riesz Regression approach. Confidence intervals are Wald-type and calculated using the estimated efficient influence functions of the estimands.}
    \label{tab:est}
\end{table}

We implement our test using the RR approach introduced in the last subsection. For confidence intervals, we leverage the asymptotic normality of the estimators and estimate the variance using the mean squared estimated efficient influence functions \citep{chernozhukov_2024, liu_general_2024}. We present our estimates in Table \ref{tab:est}. Our estimate of $\text{TE} -\text{TE}^{R}$ is significantly different from 0. Therefore, we reject the null hypothesis that $\text{NIE} = \text{NIE}^{R}$ and $\text{NDE} = \text{NDE}^{R}$. In this empirical example, one should not interpret the RIA estimates as the natural effects. Furthermore, the sum of the absolute differences between the NIE and the NIE$^{R}$ and between the NDE and the NDE$^{R}$ is as large or larger than $\vert \text{TE} -\text{TE}^{R} \vert$, which is estimated to be 0.0101. 

\section{Covariance perspective}
Next, we characterize the differences between natural effects and their RIAs analytically and provide substantive intuition using a covariance-based representation. 
For expositional clarity, we first focus on a scalar binary mediator and no baseline confounder $C$. This simple case most clearly captures our core intuition. Next, we generalize the covariance representation to vector-valued mediators with arbitrary distributions and baseline confounders. The expressions are derived using only the definitions of the estimands, without imposing any identifying assumptions or functional form restrictions. 

\subsection{Single binary mediator, no baseline confounders}
\label{sec: cov_binary}

We first provide succinct expressions for the NIE and NIE$^R$.

\begin{lemm}
When $C=\varnothing$, and the support of $M$ is $\{0,1\}$,
    $\text{NIE}=\E[(M_1-M_0)(Y_{1,1}-Y_{1,0})]$, and $\text{NIE}^R = \E(M_1-M_0)\E(Y_{1,1}-Y_{1,0})$. \label{lemma:binary}
\end{lemm}

Lemma \ref{lemma:binary} shows that the NIE is the expectation of a product and the NIE$^R$ is the product of expectations. Thus, their difference is the covariance between the two effects. Similarly, we can also express the difference between the NDE and the NDE$^R$ as a covariance. 

\begin{prop}
When $C=\varnothing$, and the support of $M$ is $\{0,1\}$, \label{prop:binary}
\begin{align*}
    \text{NIE} - \text{NIE}^R &= \Cov(M_1 - M_0, Y_{1,1}-Y_{1,0}) \\
    \text{NDE} -\text{NDE}^R &= \Cov(M_0, Y_{1,1}-Y_{1,0}-Y_{0,1}+Y_{0,0}).
\end{align*}
\end{prop}

Generally, the natural effects and the RIAs differ to the extent that the potential mediators ($M_a$) and the potential outcomes ($Y_{a',m}$) are correlated with each other. This makes sense as the RIAs are defined using random draws of potential mediators, $G_a$, that are independent of $Y_{a',m}$, whereas the natural effects do not remove the naturally occurring dependency between the potential mediators and the potential outcomes. 


The difference between the NIE and the NIE$^R$ equals the covariance between the treatment effect on the mediator ($M_1-M_0$) and the mediator effect on the outcome when treatment is set to 1 ($Y_{1,1}-Y_{1,0}$). Since Proposition \ref{prop:binary} assumes the absence of pre-treatment confounders, $C=\varnothing$, the source of this covariance must be a post-treatment confounder, $L$. Thus, the NIE and NIE$^R$ will differ if any $L$ modifies both the effect of the treatment on the mediator and the effect of the mediator on the outcome, which we call shared modification. Conversely, the NIE and NIE$^R$ will coincide in the absence of shared modification. 

We note that it is irrelevant whether there is a treatment-induced confounder. What matters for the equality of NIE and NIE$^R$ is whether there is a shared modifier. The NIE and the NIE$^R$ may coincide in the presence of treatment-induced confounding, and they may differ in its absence. Finally, the equality of NIE and NIE$^R$ ensures the identification of both NIE and NDE under the weak assumptions for the RIAs.\footnote{The weak assumptions for the RIAs identify both the NIE$R$ and ATE. Since ATE=NIE+NDE, as long as the ATE and NIE are both identified, the NDE is also identified.}

In the MTO example, the effect of voucher assignment ($A$) could be modified by voucher take-up ($L$), while voucher take-up may increase the effect of moving to a lower-poverty neighborhood ($M$) by reducing the cost associated with the latter. 
In that case, the covariance between the treatment effect on the mediator and the mediator effect on the outcome will be positive.\footnote{The fact that $\text{NIE}^R = \E(M_1-M_0)\E(Y_{1,1}-Y_{1,0})$ attests that NIE$^R$ is in fact aligned with the traditional product method of estimating direct effects \citep{baron_moderatormediator_1986}, in the sense that it is the product of two average effects. \citet[][p.260]{glynn_product_2012} discusses the fallacy of using the product method to estimate the NIE. However, unlike our nonparametric analysis, \citeauthor{glynn_product_2012}'s (2012) results are highly parametric and hence less general.}

The difference between the NDE and the NDE$^R$ is the covariance between the mediator value under control ($M_0$) and the interaction effect between the treatment and the mediator on the outcome ($Y_{1,1}-Y_{1,0}-Y_{0,1}+Y_{0,0}$). For this covariance to be non-zero, there first needs to be variation in the treatment-mediator interaction effect. Furthermore, this variation has to co-vary with $M_0$. Since all treatment-mediator confounders are, by construction, included in $L$, such co-variation exists to the extent that $L$ modifies the treatment-mediator interaction effect. 

In our empirical example, it seems implausible that the assignment of the voucher would change the effect of living in a low-poverty neighborhood on mental health, hence, we might theoretically rule out deviation of the NDE$^R$ from the NDE. In the setting considered by \citet{zhou_attendance_2022}, college attendance is the treatment, and college graduation is the mediator. In that case, $M_0$ is necessarily 0 for everyone (if someone does not attend college, they cannot graduate). Consequently, we can also rule out deviation of the NDE$^R$ from the NDE. In both these settings, the estimated $\text{TE} - \text{TE}^R$ would capture $\text{NIE} - \text{NIE}^R$ alone, making our empirical test in Section \ref{sec: test} solely a test of $\text{NIE} = \text{NIE}^R$.

\citet{miles_causal_2023} proposes a set of mediation null criteria. In particular, the definition of his ``sharper mediation null'' condition is: For each individual in the population, either $M_1=M_{0}$ or $Y_{a,m}=Y_{a,m'}$ for all
$a$, $m$, and $m'$. The corresponding null criterion states that a valid measure of indirect effect should be zero when the sharper mediation null condition is true. By Lemma \ref{lemma:binary}, the NIE clearly satisfies this criterion, while the NIE$^R$ does not. For example, if half of the population has $M_1 - M_0=1$ and $Y_{1,1}-Y_{1,0}=0$ while the other half has $M_1 - M_0=0$ and $Y_{1,1}-Y_{1,0}=1$, the NIE will be zero, but the NIE$^R$ will be $1/4$. 

In terms of the NIE and the NIE$^R$, Proposition \ref{prop:binary} expands on and demystifies \citet{miles_causal_2023} in two ways. First, the null condition is arguably a knife-edge scenario. Our result, in contrast, provides a complete characterization of the difference between NIE and the NIE$^R$, regardless of whether the null condition holds. Second, \citet{miles_causal_2023} proves that the NIE$^R$ does not satisfy the null criterion using a specific numerical counterexample, which might be viewed as a contrived example \citep[][p.1163]{miles_causal_2023}. By contrast, Proposition \ref{prop:binary} analytically reveals why and when the NIE$^R$ deviates from the null criterion: it is because the NIE$^R$ omits the natural dependency between the treatment effect on the mediator and the mediator effect on the outcome, which happens when some post-treatment confounders modify the mediator effect on the outcome.
To the extent that this is common in practice, there is nothing ``contrived'' in the NIE$^R$'s violation of the null criterion. 

\subsection{General case}

In last subsection, we focused on the case of a binary $M$ and no baseline confounder $C$. Now we generalize our results to the case where there are arbitrary vectors of mediators and baseline confounders. Again, we do not impose any identifying assumptions or parametric restrictions.

\begin{prop} 
\phantom{.} \label{prop: general}
\begin{align}
    \text{NIE} - \text{NIE}^{R} &= \sum_{m \in \mathcal{M}} \E\{\Cov[\one(M_1=m)-\one(M_0=m), Y_{1,m} \mid C] \} \nonumber \\
    \text{NDE} - \text{NDE}^{R} &= \sum_{m \in \mathcal{M}} \E\{\Cov[\one(M_0=m), Y_{1,m}-Y_{0,m} \mid C] \}, \nonumber
\end{align}
where $\one(\cdot)$ is the indicator function, and $\mathcal{M}$ is the support of $M$. The relationships above directly hold for multivalued discrete mediators, but they also hold for continuous mediators if summations are replaced with integrals and the indicator function is replaced with the Dirac delta function. 
\end{prop}

We thus obtain a covariance-based representation analogous to Proposition \ref{prop:binary}. Here, the building blocks are conditional covariances between the potential mediators ($M_a$) and the potential outcomes ($Y_{a',m}$) given baseline confounders $C$. We further summarize the $c$- and $m$-specific covariances by taking expectation over the distribution of $C$ and  taking sum over the support of $M$. Again, the natural effects and the RIAs generally differ due to the dependency between the mediator and outcome potential values conditional on baseline confounders. Clearly, the natural effects and the RIAs coincide when the cross-world independence assumption (Assumption \ref{assu: cross}) is satisfied.


An alternative RIA-based decomposition is developed by \citet{lok_defining_2016} and \citet{lok_causal_2021}
In this decomposition, the TE is decomposed to what are called the organic indirect and direct effects ($\text{NIE}^{\text{organic}}$ and $\text{NDE}^{\text{organic}}$). 
\begin{equation*}
    \underbrace{\E(Y_1-Y_0)}_{\text{TE}}= \underbrace{\E(Y_{1}-Y_{1,G_{0}})}_{\text{NIE}^{\text{organic}}}  + \underbrace{\E(Y_{1,G_{0}}-Y_{0})}_{\text{NDE}^{\text{organic}}}.
\end{equation*}

We again show a corresponding covariance representation in the general case.\footnote{ \citet{zheng_longitudinal_2017} propose another related decomposition \citep[also see][p.264]{nguyen_clarifying_2022}. The intervention underlying this decomposition involves assigning to people with $C=c, L_a=l$ values of mediator randomly drawn from the distribution of $M_{a'}$ conditional on $C=c, L_{a'}=l$. The differences between the natural effects and components of this decomposition do not have a covariance representation. This is because the way $L$ enters into the NIE's counterpart in this decomposition makes it the path-specific effect through $M$ but not $L$ (see Appendix S2 in \citeauthor{miles_causal_2023} [2003] and Appendix S8 in \citeauthor{diaz_nonparametric_2021} [2021]). Thus, the components of this decomposition are conceptually further removed from the natural effects.}
\begin{prop}
\begin{align*}
    \text{NIE} - \text{NIE}^{\text{organic}} &= -\sum_{m \in \mathcal{M}} \E\{\Cov[\one(M_0=m), Y_{1,m} \mid C] \} \\
    \text{NDE} - \text{NDE}^{\text{organic}} &= \sum_{m \in \mathcal{M}} \E\{\Cov[ \one(M_0=m), Y_{1,m}\mid C] \}.
\end{align*}
\end{prop}


\section{Structural equation perspective}
In this section, we illustrate some specific data generating processes (DGPs) that would make the NDE coincide with the NDE$^R$ or the NIE with the NIE$^R$. We have provided a covariance-based representation of the differences between the natural effects and the RIAs, now we further develop tools for substantively reasoning when the covariances would disappear. 
We express these DGPs using structural equations (generative models) with parametric constraints. Throughout this section, we do not restrict the dimension or the distribution of mediators. 

We first present results with assumed linearity and without baseline confounders, which provides the easiest intuition. Then we extend the results to structural equations without the linearity restrictions and treatment randomization. For comparison with parametric constraints below, we note that the nonparametric structural equations with no constraints are as follows:
    \begin{align*}
    C &= g_C(\epsilon_C) \\
    A &= g_A(C, \epsilon_A) \\
        L &= g_L(C,A, \epsilon_L)  \\
        M &= g_M(C,A,L, \epsilon_M) \\
        Y &= g_Y(C,A,L,M, \epsilon_Y), 
    \end{align*} 
where $g_C$, $g_A$, $g_L$, and $g_M$ are arbitrary functions of their arguments. And $\epsilon_C, \epsilon_A, \epsilon_L, \epsilon_M$ and $\epsilon_Y$ are unspecified inputs for each variable. Importantly, throughout this section, we allow these unspecified inputs to be arbitrarily dependent on one another and all specified variables. This makes our setting more general then the nonparametric structural equations that are commonly represented by directed acyclic graphs \citep{pearl_causal_1995,pearl_causal_2012}. 

\subsection{Linear structural equations, no baseline confounders}
Since $C$ is empty, we consider the structural equations for $A$, $L$, $M$, and $Y$. In this subsection, the notation technically only applies to one $L$ and one $M$ variables, but our expressions can be easily extended to accommodate multiple $L$ and $M$ variables without compromising the intuition.
\begin{prop}
    Under the following linear structural equations with constant coefficients (i.e., all $\alpha$, $\beta$, $\gamma$ terms are constants),\label{prop: linear}
\begin{align*}
    A &= \epsilon_A \\
    L &= \alpha_0 + \alpha_1 A + \epsilon_L \\
    M &= \beta_0 + \beta_1 A + \beta_2 L + \beta_3AL + \epsilon_M \\
    Y &= \gamma_0 + \gamma_1 A + \gamma_2 L + \gamma_3 M + \gamma_4 AL + \gamma_5 AM + \gamma_6 LM + \gamma_7 ALM + \epsilon_Y,   
\end{align*}
we have $\text{NIE}-\text{NIE}^R = (\gamma_6 +\gamma_7)\beta_3 \Var(\epsilon_L)$, and $\text{NDE}-\text{NDE}^R = \gamma_7\beta_2 \Var(\epsilon_L) + \gamma_7 \Cov(\epsilon_L, \epsilon_M)$.\footnote{Clearly, Proposition \ref{prop: linear} is a special case of Proposition \ref{prop: general}. Additionally, when $M$ is binary, Proposition \ref{prop: linear} is also a special case of Proposition \ref{prop:binary}.}
\end{prop}
Hence, under the linear structural equations, there are multiple sufficient conditions for either the NIE or the NDE to coincide with their respective RIAs. The NIE and the NIE$^R$ are equivalent if 1) there is no $AL$ interaction in the equation for $M$, i.e., $\beta_3 = 0$; or 2) if there is no $LM$ interaction in the equation for $Y$, i.e., $\gamma_6=\gamma_7=0$. These results are consistent with Proposition \ref{prop:binary}, which showed that shared modification of the effect of $A$ on $M$ and the effect of $M$ on $Y$ by $L$ causes the deviation between NIE and NIE$^R$ when $M$ is binary. Here, the effect of $A$ on $M$ is modified by $L$ if and only if $\beta_3 \neq 0$, and the effect of $M$ on $Y$ is modified by $L$ if and only if $\gamma_6 \neq 0$ or $\gamma_7 \neq 0$.

The NDE and the NDE$^R$ are equivalent if 1) there is no three-way interaction $ALM$ in the equation for $Y$, i.e., $\gamma_7 = 0$; or 2) $L$ does not have an effect on $M$ when $A=0$, and there is no unaccounted common determinants of $M$ and $L$, i.e, $\beta_2=0$ and $\Cov(\epsilon_L, \epsilon_M)=0$. Recall that in Proposition \ref{prop:binary}, we showed that, when $M$ is binary, NDE and the NDE$^R$ differ to the extent that $L$ modifies the treatment-mediator interaction effect on $Y$. This modification takes the form of the three-way interaction $ALM$ in Proposition \ref{prop: linear}. 

In summary, equivalences can be established by ruling out certain interaction effects. It is possible to have only one of the NIE and the NDE coincide with their RIA. When only one of the natural effects equal its RIA, our test parameter in Section \ref{sec: test}, $\text{TE}-\text{TE}^R$, will capture the deviation of the other natural effect from its RIA. Next subsection shows that the intuitions from the linear analysis can be extended to the settings where the structural equations are much more unrestricted.  

\subsection{Nonlinear structural equations with baseline confounders}
Throughout this subsection, we focus on constraints on the structural equations for $Y$. Thus, we maintain completely unconstrained structural equations for $C$, $A$, $L$, and $M$. Below, we let $g_{Y1}$ and $g_{Y2}$ denote arbitrary functions of their arguments. Thus, within these functions, the effects of the variables are left completely unconstrained. 

\begin{prop}
If $Y = g_{Y1}(C,A,L,\epsilon_{Y1}) + g_{Y2}(C,M,\epsilon_{Y2})$, NDE=NDE$^R$. \label{prop:NDE}
\end{prop}
The structural equation rules out $AM$ and $LM$ interactions in the equation for $Y$, in the sense that the nonparametric function containing $M$ is additively separable from the nonparametric function containing $A$ and $L$. We only focus on the direct effects in this proposition because the analogous condition for the indirect effects is too strong (see Appendix A4). 

In summary, in the presence of treatment-induced confounders, it is still possible that $\text{NIE}=\text{NIE}^R$ or $\text{NDE}=\text{NDE}^R$. However, these equivalences require imposing constrains on  relevant structural equations by ruling out interaction effects. The structural equation constraints we present are sufficient but not necessary to establish equivalences between the natural effects and the RIA. Nevertheless, they are derived with the goal of being maximally flexible, in the sense that they allow as much complexity in functional form as possible without incurring other strong constraints.

\section{Related estimands}
The theory we developed for causal mediation analysis proves to be useful for unifying three long-standing literatures in causal inference. In causal inference, it is not unusual that a pair of competing estimands is present, where one has a more natural interpretation and the other is easier to identify. Apart from the natural mediation effects and their RIAs, we discuss two other such pairs of estimands: the average treatment effect (ATE) versus the local average treatment effect (LATE) in the IV context \citep{angrist_identification_1996}; and what we call the natural Mann-Whitney estimand and its RIA \citep{mann_whitney_test_1947}. Specifically, we establish a formal equivalence result between estimands in the IV literature and the mediation literature. And we reveal a striking resemblance between the Mann-Whitney estimands and the mediation estimands. 

\subsection{ATE and LATE}
We first define the ATE and LATE estimands. In keeping with the notation we used for causal mediation analysis above, we consider three temporally ordered variables, $A$, $M$, and $Y$. In the IV context, $A$ is the IV, $M$ is the treatment, and $Y$ is the outcome. Here, we focus on the case where $A$ and $M$ are both binary, and $A$ is randomized, which is a classic setting considered in the IV literature \citep{angrist_identification_1996,balke_bounds_1997}. Then, the ATE is defined as $\E(Y_{M=1}-Y_{M=0})$, and the LATE is defined as $\E(Y_{M=1}-Y_{M=0} \mid M_{A=1}=1, M_{A=0}=0)$, i.e., the average effect of $M$ on $Y$ among those whose $M$ value is induced to increase by an increase in $A$ (those who are the ``compliers''). In this subsection, we explicitly write the assignment variables in the potential outcomes to avoid ambiguity. 
Also note that the labelling of the ``treatment'' variable differs between the IV and mediation contexts: in the IV context, the treatment refers to $M$, while in the mediation context, it refers to $A$.

In the IV context, the estimand with a more natural interpretation is the ATE, while the LATE requires weaker identifying assumptions \citep{robins_identification_1996,Imbens_2010, aronow_beyond_2013, wang_bounded_2018}. Just like in the mediation context, applied researchers often interpret a LATE estimate as if it was the ATE \citep{aronow_beyond_2013, sarvet_interpretational_2023}. We show that there exists a direct equivalence between $\text{ATE}-\text{LATE}$ and $\text{NIE}-\text{NIE}^R$ under four standard identifying assumptions for the LATE: 1) Exclusion: $Y_{A=a, M=m}=Y_{M=m}, \forall \{a,m\}$; 2) Independence: $A \indep \{M_{A=1}, M_{A=0}, Y_{A=1}, Y_{A=0}\}$; 3) Relevance: $\E(M \mid A=1)-\E(M \mid A=0) >0$; and 4) Monotonicity: $M_{A=1} \geq M_{A=0}$ almost surely. We also denote the identified functional called the Wald estimand as $\text{Wald} \defeq \frac{\E(Y \mid A=1)-\E(Y \mid A=0)}{\E(M \mid A=1)-\E(M \mid A=0)}$.

\begin{prop}
    Under the assumptions of exclusion, independence, and relevance, $$\text{Wald}-\text{ATE}=\frac{\Cov(M_{A=1}-M_{A=0}, Y_{M=1}-Y_{M=0})}{\E(M_{A=1}-M_{A=0})}=\frac{\text{NIE}-\text{NIE}^R}{\E(M_{A=1}-M_{A=0})},$$ 
    which, further under monotonicity, also equals $\text{LATE}-\text{ATE}$. Here, NIE$^R$ is defined with $C=\varnothing$. \footnote{Also, by Proposition \ref{prop:binary} and the exclusion assumption, $\text{NIE}-\text{NIE}^R=\text{TE}-\text{TE}^R$.} \label{prop:IV}
\end{prop}

Thus, under the four assumptions identifying the LATE, the difference between the LATE and the ATE is simply the difference between the NIE and the NIE$^R$ scaled by the effect of $A$ on $M$. This means that, under these assumptions, the LATE differs from the ATE if and only if the NIE differs from the NIE$^R$. For intuition on $\text{LATE}-\text{ATE}$, notice that $\Cov(M_{A=1}-M_{A=0}, Y_{M=1}-Y_{M=0})=\Cov[\one(M_{A=1}=1, M_{A=0}=0),Y_{M=1}-Y_{M=0}]$ captures selection into the subpopulation of compliers based on the effect of $M$ on $Y$. If there is strong selection, then the local average effect of $M$ on $Y$ among compliers must differ substantially from the corresponding global average effect.

There is a long-standing literature on using the Wald estimand to estimate the ATE based on exclusion, independence, relevance, and another additional assumption \citep{heckman_instrumental_1997,hernan_instruments_2006,wang_bounded_2018}.
A weak form of the additional assumption has recently appeared in \citeauthor{hernan_causal_2020} (2020, Section 16.3) and \citet{hartwig_average_2023}, which can be written as $\Cov(M_{A=1}-M_{A=0}, Y_{M=1}-Y_{M=0})=0$. Proposition \ref{prop:IV} shows that this is, in fact, the weakest possible among such assumptions.

\subsection{Natural Mann-Whitney estimand and its RIA}
We define the natural Mann-Whitney estimand as $\E[\one(Y_1 \geq Y_0)]$, i.e., the probability of the potential outcome under treatment being greater than or equal to the potential outcome under control. It is often referred to as the probability of no harm (the probability of the treatment not worsening the outcome), given that a larger value of $Y$ is desired. This estimand is broadly useful for scale-free evaluation of treatment effects, especially for ordinal outcomes.\footnote{A related estimand, $\Pro(Y_1 > Y_0 \mid A=1)/\Pro(Y_1=1  \mid A=1)$, for a binary $Y$, is called the probability of necessity \citep{tian_probabilities_2000}.} We call this estimand a ``natural'' estimand, because it is an aggregation of an individual-level contrast of potential outcomes. 

The natural Mann-Whitney estimand is difficult to identify for the same reason that the NIE and the NDE are difficult to identify: just like $\E(Y_{1, M_0})$, the natural Mann-Whitney estimand (non-linearly) involves the assignment of two different treatment values to the same individual. Due to the fundamental problem of causal inference \citep{holland_1986}, the joint distribution of two potential outcomes is impossible to nonparametrically identify even with a randomized treatment.\footnote{By contrast, the TE is a linear combination of two treatment values, avoiding the cross-world assignment problem simply due to the equality $\E(Y_1-Y_0)=\E(Y_1)-\E(Y_0)$.} Hence, an assumption analogous to cross-world independence (Assumption \ref{assu: cross}) can also be used to identify the natural Mann-Whitney estimand: $Y_1 \indep Y_0$ \citep{greenland_causal_2020}, which can be relaxed to a conditional version: $Y_1 \indep Y_0 \mid C$. However, even the conditional version of this assumption is unlikely to hold, because it requires that all variables affecting $Y$ under both treatment and control are measured.\footnote{Assumptions of the same form are also invoked to identify principal stratum estimands in clinical trial contexts \citep{hayden_estimator_2005, qu_general_2020}, which is a practice extensively criticized by \citet{vansteelandt_chasing_2024}.} 

Consequently, an alternative estimand has been used in practice: $\E[\one(H_1 \geq H_0)]$, where $H_a$ is a value randomly drawn from the marginal distribution of $Y_a$. Clearly, this alternative estimand has the interpretation of a RIA. In contrast to the natural Mann-Whitney estimand, the Mann-Whitney RIA does not aggregate an individual-level contrast. On the other hand, randomization of treatment does enable the identification of the Mann-Whitney RIA. The Mann-Whitney RIA has a long history in statistics, dating back to the Mann-Whitney $U$ test \citep{mann_whitney_test_1947} and the Wilcoxon rank-sum test \citep{wilcoxon_individual_1945}. Recent methodological developments based on the Mann-Whitney RIA include the probability index model \citep{thas_probabilistic_2012}, the win ratio \citep{pocock2012}, the efficient estimation of the RIA \citep{mao_causal_2018}, a local version of the RIA in the presence of noncompliance \citep{mao_wilcoxon-mann-whitney_2024}, and the rank average treatment effect \citep{lei_causal_2024}. 

Similar to the mediation literature, conflation of the natural Mann-Whitney estimand and its RIA is pervasive even in methodological work. For example, in a textbook discussion on the Mann-Whitney RIA, \citet{thas_comparing_2010} claims that ``If this conclusion is statistically significant, it is very relevant evidence to a physician that most of his patients will be better
off with the treatment.'' \citet{wu_causal_2014} states ``This allows us to make inference about the potential outcome-based $\delta$ through the estimable quantity $\xi$...", where $\delta$ and $\xi$ are respectively the natural Mann-Whitney estimand and its RIA. And \citet{demidenko_p_2016} names the Mann-Whitney RIA the ``$D$-value'' and argues that ``The $D$-value has a clear interpretation as the proportion of patients who get worse after the treatment'', in the context where a smaller value of a continuous $Y$ is desirable. 

Interestingly, despite (or maybe due to) recurrent confusion, the literature on Mann-Whitney estimands has been clarifying the important differences between the natural Mann-Whitney estimand and its RIA since decades before \citet{miles_causal_2023} pioneered an analogous inquiry in causal mediation analysis. The early work of \cite{hand_comparing_1992} already notes the possibility of sign reversal in the relationship between the natural Mann-Whitney estimand and its RIA (when $1/2$ is subtracted from both), which has been known as Hand's paradox. Multiple works since have considered various DGPs under which Hand's paradox is present or absent \citep{hand_comparing_1992, fay_causal_2018, greenland_causal_2020}. This line of work is in the same spirit as our theoretical analysis on the relationship between the natural mediation estimands and their RIAs.

Lastly, there is also a covariance representation for the difference between the natural Mann-Whitney estimand and its RIA. 
\begin{prop}
\begin{align*}
    \E[\one(Y_1 \geq Y_0)] - \E[\one(H_1 \geq H_0)] = \sum_{t \in \mathcal{T}}\sum_{s \in \mathcal{S}} \one(t \geq s) \Cov[\one(Y_1 = t),\one(Y_0=s)],
\end{align*}
where $\mathcal{T}$ and $\mathcal{S}$ are respectively the supports of $Y_1$ and $Y_0$.
When $Y$ is binary with the support of $\{0,1 \}$, the expression simplifies to $\Cov(Y_1, Y_0)$.
\end{prop}
Clearly, the natural Mann-Whitney estimand differs from its RIA to the extent that $Y_1$ and $Y_0$ are dependent on each other. This is in parallel to the natural mediation effects differing from their RIAs to the extent that $M_a$ and $Y_{a',m}$ are dependent. By redefining the estimands using random draws, RIAs in both cases miss a naturally occurring dependency. The thorny issue created by cross-world treatment assignments for identification cannot be magically waved away by redefining the estimand.


\section{Conclusion}
In this paper, we answer the question of when natural mediation estimands coincide with or differ from their randomized interventional analogues. In order to do so, we provide tools for both empirical testing and theoretical reasoning to researchers who wish to estimate and interpret the RIAs. 
Our test and theories are complementary to one another: when the researcher empirically rejects the null hypothesis of the test, they can conclude with confidence (up to the chosen significance level) that the natural effects and the RIAs are \emph{different}; when the researcher has theoretical support for specific structural equations, they may reasonably posit that a particular natural effect and its corresponding RIA are \emph{equivalent}. With respect to the two theoretical perspectives, the covariance perspective is complete, in the sense that it provides necessary and sufficient conditions for the equivalence between the natural effects and the RIAs; while the structural equation perspective provides simple and intuitive sufficient conditions of equivalence even when $M$ is vector-valued with arbitrary distributions.

A common dilemma facing researchers across three fields of causal inference (causal mediation analysis, instrumental variable, and Mann-Whitney estimands) is that a natural estimand is more interpretively appealing but hard to identify while an alternative estimand is less appealing but easier to identify. Going forward, we recommend four strategies to applied researchers in all three areas. First, we join \citet{sarvet_interpretational_2023} to call for more precision in interpreting estimates of the alternative estimands. Second, with the addition of our two theoretical perspectives in this paper, now researchers in all three areas are able to reason about when the natural estimand coincides with, or at least does not have the opposite sign to, the alternative estimand. Third, in all three areas, bounding methods have been developed to provide partial identification for the natural estimands \citep[e.g.,][]{miles_partial_2017,swanson_partial_2018, lu2020}. Fourth, in causal mediation analysis, we uniquely provide a falsification test for interpreting the RIAs as the natural mediation effects, which goes beyond theoretical reasoning and provides empirical guidance. 

\section*{Acknowledgement}
We are grateful for a comment from a reviewer at the Annals of Applied Statistics for \citet{yu2025nonparametric}, which gave us the initial inspiration. We also thank  Sameer Deshpande, Eric Grodsky, Nick Mark, Xinran Miao, Chan Park, Michael Sobel, Jiwei Zhao, Xiang Zhou, and especially Hyunseung Kang for helpful suggestions. for helpful suggestions. An earlier version of this paper was presented at the American Causal Inference Conference and the International Biometric Conference. We thank the audience at these conferences for engaging discussions. The first author's travel to ACIC was funded by the Archie O. Haller Memorial Scholarship from UW–Madison’s Department of
Sociology.

\section*{Appendices}
\subsection*{A1. Proof of Proposition 1}
The NIE and NDE are defined in terms of $\E(Y_{a,M_{a'}})$  for two treatment values ($a,a'$).  When $M$ is binary and its support is $\{0,1\}$, we rewrite this quantity just using its definition:
\begin{align*}
    &\phantom{{}={}} \E(Y_{a,M_{a'}}) \\
    &= \E[Y_{a,1}M_{a'} + Y_{a,0}(1-M_{a'})] \\
    &= \E(Y_{a, 0}) + \E[ M_{a'}(Y_{a, 1}-Y_{a, 0}) ] \\
    &= \E(Y_{a, 0}) + \E\{ \E[ M_{a'}(Y_{a, 1}-Y_{a, 0}) \mid C] \}.
\end{align*}

The NIE$^R$ and NDE$^R$ are defined in terms of $\E(Y_{a, G_{a'}})$ for two treatment values ($a,a'$). When $M$ is binary, we again rewrite this quantity using its definition:
\begin{align*}
   &\phantom{{}={}} \E(Y_{a, G_{a'}}) \\
   &=  \E[\E(Y_{a, G_{a'}} \mid C)] \\
   &=\E[\E(Y_{a, 1} \mid G_{a'}=1, C)\Pro(G_{a'}=1 \mid C) + \E(Y_{a, 0} \mid G_{a'}=0, C)\Pro(G_{a'}=0 \mid C) ]\\
   &=\E\{\E(Y_{a, 1} \mid C)\E(M_{a'} \mid C) + \E(Y_{a, 0} \mid C)[1-\E(M_{a'} \mid C) ] \} \\
   &= \E(Y_{a, 0}) + \E\{\E(M_{a'} \mid C) [\E(Y_{a, 1} - Y_{a, 0} \mid C)] \} \\
   &=\E(Y_{a,M_{a'}}) - \E[\Cov(M_{a'}, Y_{a, 1}-Y_{a, 0} \mid C)].
\end{align*}

Then using the results above, we have the following representations:
\begin{align*}
    \text{NIE} &= \E(Y_{1,M_1}-Y_{1,M_0}) = \E[(M_1 - M_0)(Y_{1,1}-Y_{1,0})] \\
    \text{NIE}^R &= \E(Y_{1,G_1}-Y_{1,G_0}) = \E[\E(M_1 - M_0 \mid C) \E(Y_{1,1}-Y_{1,0} \mid C) ] \\
    \text{NDE} &= \E(Y_{1,M_0}-Y_{0,M_0}) = \E(Y_{1,0}-Y_{0,0}) + \E\{ M_0 [Y_{1,1}-Y_{1,0}-(Y_{0,1}-Y_{0,0})] \} \\
    \text{NDE}^R &= \E(Y_{1,G_0}-Y_{0,G_0}) = \E(Y_{1,0}-Y_{0,0}) + \E\{ \E(M_0 \mid C) \E [Y_{1,1}-Y_{1,0}-(Y_{0,1}-Y_{0,0}) \mid C] \}.
\end{align*}

Hence,
\begin{align*}
    \text{NIE} &= \text{NIE}^R + \E[\Cov(M_1 - M_0, Y_{1,1}-Y_{1,0} \mid C)] \\
    \text{NDE} &= \text{NDE}^R + \E\{\Cov[M_0, Y_{1,1}-Y_{1,0}-(Y_{0,1}-Y_{0,0}) \mid C] \}\\
    \text{TE} &= \text{TE}^R + \E[\Cov(M_1, Y_{1,1}-Y_{1,0} \mid C) - \Cov(M_0, Y_{0,1}-Y_{0,0} \mid C)].
\end{align*}
When $C$ is an empty set, we obtain the results shown in Proposition 1.

\subsection*{A2. Proof of Propositions 2 and 3}
The NIE and NDE are still defined in terms of $\E(Y_{a,M_{a'}})$  for two treatment values ($a,a'$). Treating $M$ as a vector of continuous variables, we rewrite this quantity using its definition:
\begin{align*}
    &\phantom{{}={}} \E(Y_{a,M_{a'}}) \\
    &=\E\left[ \int Y_{a,m} \one(M_{a'}=m)  \dd m \right] \\
    &= \int \E[Y_{a,m} \one(M_{a'}=m)] \dd m \\
    &= \int \E\{\E[Y_{a,m} \one(M_{a'}=m) \mid C] \} \dd m,
\end{align*}
where the first equality holds by treating the Dirac delta function $\one(M_{a'}=m)$ as a limiting case of a probability density function concentrated at $M_{a'}=m$. This allows us to express a function of $M_{a'}$ as an integral over the support of $M_{a'}$.

The NIE$^R$ and NDE$^R$ are defined in terms of $\E(Y_{a, G_{a'}})$ for two treatment values ($a,a'$). We rewrite this quantity as follows:
\begin{align*}
   &\phantom{{}={}} \E(Y_{a, G_{a'}}) \\
   &= \E[\E(Y_{a, G_{a'}} \mid C)] \\
   &=\iint \E(Y_{a, m} \mid G_{a'}=m, C=c) f_{G_{a'} \mid c}(m) f_C(c) \dd m \dd c \\
   &=\iint \E(Y_{a, m} \mid C=c ) f_{M_{a'} \mid c}(m) f_C(c) \dd m \dd c \\
   &=\iint \E(Y_{a, m} \mid C=c ) \E[\one(M_{a'}=m) \mid C=c] f_C(c) \dd m \dd c, 
\end{align*}
where the last equality is by the property of the Dirac delta function $\one(M_{a'}=m)$.

Therefore,
\begin{align*}
    \text{NIE} &= \E(Y_{1,M_1}-Y_{1,M_0}) = \int \E\{ \E\{ [\one(M_1=m)-\one(M_0=m)] Y_{1,m} \} \mid C \} \dd m \\
    \text{NIE}^R &= \E(Y_{1,G_1}-Y_{1,G_0}) = \int \E\{\E [\one(M_1=m)-\one(M_0=m) \mid C] \E( Y_{1,m} \mid C ) \} \dd m \\
    \text{NDE} &= \E(Y_{1,M_0}-Y_{0,M_0}) = \int \E\{\E[ (Y_{1,m}-Y_{0,m})\one(M_0=m) \mid C ]\} \dd m \\
    \text{NDE}^R &= \E(Y_{1,G_0}-Y_{0,G_0}) = \int \E\{ \E[ (Y_{1,m}-Y_{0,m}) \mid C]\E[\one(M_0=m) \mid C] \} \dd m.
\end{align*}

And
\begin{align*}
    \text{NIE} &= \text{NIE}^R + \int \E\{\Cov[\one(M_1=m)-\one(M_0=m), Y_{1,m} \mid C]\} \dd m \\
    \text{NDE} &= \text{NDE}^R + \int \E\{\Cov[Y_{1,m}-Y_{0,m}, \one(M_0=m) \mid C] \} \dd m \\
    \text{TE} &= \text{TE}^R + \int \E\{\Cov[Y_{1,m}, \one(M_1=m) \mid C]\} - \E\{\Cov[Y_{0,m}, \one(M_0=m) \mid C] \}\dd m.
\end{align*}
When $M$ is a vector of discrete variables, we replace the integrals with summations to obtain the results in Proposition \ref{prop: general}.

Proposition 3 similarly follows from the expressions of $\E(Y_{a,M_{a'}})$ and $\E(Y_{a,G_{a'}})$ derived above.

\subsection*{A3. Proof of Proposition 4}
We let $L_a$ denote the potential values of $L$ under treatment assignment $a$.
Under the structural equations of Proposition 4,
\begin{align*}
    Y_{1 M_1} &= \gamma_0 + \gamma_1 + (\gamma_2+\gamma_4) L_1 + (\gamma_3+\gamma_5) M_1 + (\gamma_6+\gamma_7) L_1 M_1 + \epsilon_Y  \\
    Y_{1 M_0} &= \gamma_0 + \gamma_1 + (\gamma_2+\gamma_4) L_1 + (\gamma_3+\gamma_5) M_0 + (\gamma_6+\gamma_7) L_1 M_0 + \epsilon_Y \\
    Y_{0 M_0} &= \gamma_0 + \gamma_2 L_0 + \gamma_3 M_0 + \gamma_6L_0 M_0 + \epsilon_Y \\
    Y_{1 G_1} &= \gamma_0 + \gamma_1 + (\gamma_2+\gamma_4) L_1 + (\gamma_3+\gamma_5) G_1 + (\gamma_6+\gamma_7) L_1 G_1 + \epsilon_Y  \\
    Y_{1 G_0} &= \gamma_0 + \gamma_1 + (\gamma_2+\gamma_4) L_1 + (\gamma_3+\gamma_5) G_0 + (\gamma_6+\gamma_7) L_1 G_0 + \epsilon_Y \\
    Y_{0 G_0} &= \gamma_0 + \gamma_2 L_0 + \gamma_3 G_0 + \gamma_6L_0 G_0 + \epsilon_Y.
\end{align*}
Hence,
\begin{align*}
    \text{NDE} &= \gamma_1 + (\gamma_2 +\gamma_4)\E(L_1) -\gamma_2 \E(L_0)+ \gamma_5 \E(M_0) + (\gamma_6+\gamma_7)\E(L_1 M_0) - \gamma_6 \E(L_0M_0) \\
    \text{NDE}^R &= \gamma_1 + (\gamma_2 +\gamma_4)\E(L_1) -\gamma_2 \E(L_0)+ \gamma_5 \E(G_0) + (\gamma_6+\gamma_7)\E(L_1 G_0) - \gamma_6 \E(L_0G_0) \\
    \text{NIE} &= (\gamma_3+\gamma_5) \E(M_1-M_0) +(\gamma_6+\gamma_7)\E(L_1M_1 -L_1M_0) \\
    \text{NIE}^R &= (\gamma_3+\gamma_5) \E(G_1-G_0) + (\gamma_6+\gamma_7)\E(L_1G_1 -L_1G_0).
\end{align*}
Noting that $\E(M_a)=\E(G_a)$, and 
\begin{align*}
    &\phantom{{}={}} \E(L_a M_{a'})-\E(L_a G_{a'}) \\
    &= \E(L_a M_{a'})-\E(L_a) \E( G_{a'}) \\
    &= \Cov(L_a, M_{a'}) \\
    &= \Cov[\alpha_0 + \alpha_1 a +\epsilon_L, \beta_0+\beta_1a'+\beta_2(\alpha_0+\alpha_1a'+\epsilon_L)+\beta_3a'(\alpha_0+\alpha_1a'+\epsilon_L)+\epsilon_M] \\
    &=(\beta_2+\beta_3a')\Var(\epsilon_L) + \Cov(\epsilon_L, \epsilon_M).
\end{align*}
we have 
\begin{align*}
    \text{NDE}-\text{NDE}^R &= (\gamma_6+\gamma_7)\Cov(L_1, M_0) - \gamma_6 \Cov(L_0, M_0) \\
    &= \gamma_7\beta_2 \Var(\epsilon_L) + \gamma_7 \Cov(\epsilon_L, \epsilon_M) \\
    \text{NIE}-\text{NIE}^R &= (\gamma_6+\gamma_7)\{\Cov(L_1, M_1 ) - \Cov(L_1, M_0 ) \} \\
    &= (\gamma_6 +\gamma_7)\beta_3 \Var(\epsilon_L).
\end{align*}

\subsection*{A4. Proof of Proposition 5}
For the NDE part, our proof leverages an assumption in \citet{green_semantics_2003}: $Y_{1,m}-Y_{0,m}$ is a random variable not dependent on $m$. Originally, this assumption was proposed to identify NDE in the presence of treatment-induced confounding. We first prove that this assumption is sufficient for $\text{NDE}=\text{NDE}^R$. Then we prove that the structural equation in Proposition 5 is, in turn, sufficient for this assumption to hold. 

According to our Proposition \ref{prop: general}, we just need to show that under the assumption of \citet{green_semantics_2003}, $\int \E\{\Cov[Y_{1,m}-Y_{0,m}, \one(M_0=m) \mid C] \} \dd m=0$.
Let $Y_{1,m}-Y_{0,m}=B$, then, 
\begin{align*}
    &\phantom{{}={}} \int \E\{\Cov[\one(M_0=m), Y_{1,m}-Y_{0,m} \mid C] \} \dd m \\
    &= \int \E \{ \E[\one(M_0=m) B \mid C] - \E[\one(M_0=m) \mid C] \E(B \mid C) \} \dd m  \\
    &= \E \left\{ \E \left[ \int \one(M_0=m) \dd m B \mid C \right] - \int f_{M_0}(m \mid C) \dd m \E(B \mid C) \right\} \\
    &= \E[ \E(B \mid C)- \E(B \mid C) ]=0.
\end{align*}

Next, we show that, if $Y = g_{Y1}(C,A,L,\epsilon_{Y1}) + g_{Y2}(C,M,\epsilon_{Y2})$,  the assumption of \citet{green_semantics_2003} is satisfied. Under this structural equation for $Y$,
\begin{align*}
    &\phantom{{}={}} Y_{1,m}-Y_{0,m} \\
    &= g_{Y1}(C,1,g_L(C,1,\epsilon_{L}), \epsilon_{Y1}) + g_{Y2}(C,m,\epsilon_{Y2}) - g_{Y1}(C,0,g_L(C,0,\epsilon_{L}), \epsilon_{Y1}) - g_{Y2}(C,m,\epsilon_{Y2}) \\
    &= g_{Y1}(C,1,g_L(C,1,\epsilon_{L}), \epsilon_{Y1}) - g_{Y1}(C,0,g_L(C,0,\epsilon_{L}), \epsilon_{Y1}), 
\end{align*}
which is not dependent on $m$. 

For the NIE part, we propose a novel condition that is analogous to the assumption of \citet{green_semantics_2003} used above: $Y_{1,m}$ is a random variable not dependent on $m$. We refer to this condition as the analogous assumption. We first show that the analogous assumption is sufficient for NIE to be equal to NIE$^R$. According to Proposition \ref{prop: general}, it suffices to show $\int \E\{\Cov[\one(M_1=m)-\one(M_0=m), Y_{1,m} \mid C] \} \dd m=0$. Let $Y_{1,m}=B$, then under the analogous assumption,
\begin{align*}
    &\phantom{{}={}} \int \E\{\Cov[\one(M_1=m)-\one(M_0=m), Y_{1,m} \mid C] \} \dd m \\
    &=\int \E\{\Cov[\one(M_1=m)-\one(M_0=m), B \mid C] \} \dd m \\
    &= \int \E\{ \E[\one(M_1=m) B \mid C]-\E[\one(M_0=m) B \mid C] \\
    &\phantom{{}={}} - \E[\one(M_1=m)-\one(M_0=m) \mid C] \E(B\ \mid C) \} \dd m \\
    &= \E\{ \E[\int \one(M_1=m) \dd m B \mid C]-\E[\int \one(M_0=m) \dd m B \mid C] \\
    &\phantom{{}={}}- \E[\int \one(M_1=m)-\one(M_0=m) \dd m \mid C] \E(B\ \mid C) \} \\
    &= \E\{ \E[B \mid C] - \E[B \mid C] \} \\
    &=0.
\end{align*}

Then, we show that if $Y = (1-A)g_{Y1}(C,L,M,\epsilon_{Y1}) + Ag_{Y2}(C,L,\epsilon_{Y2})$, the analogous assumption is satisfied. Under this structural equation, $Y_{1,m}=g_{Y2}(C,g_L(C,1,\epsilon_L),\epsilon_{Y2})$, which clearly does not depend on $m$. 

Hence, a sufficient condition for NIE$=$NIE$^R$ is $Y = (1-A)g_{Y1}(C,L,M,\epsilon_{Y1}) + Ag_{Y2}(C,L,\epsilon_{Y2})$. We opt to not present this result in the main text, as this condition would in fact make both NIE and NIE$^R$ zero. This seems too strong a condition. 

\subsection*{A5. Proof of Proposition 6}
\begin{align*}
    &\phantom{{}={}} \text{Wald} \\
    &=\frac{\E(Y_{A=1}-Y_{A=0})}{\E(M_{A=1}-M_{A=0})} \\
    &=\frac{\E[(M_{A=1}-M_{A=0})(Y_{M=1}-Y_{M=0})]}{\E(M_{A=1}-M_{A=0})} \\
    &=\frac{\E(M_{A=1}-M_{A=0})\E(Y_{M=1}-Y_{M=0}) + \Cov(M_{A=1}-M_{A=0}, Y_{M=1}-Y_{M=0})}{\E(M_{A=1}-M_{A=0})} \\ 
    &= \text{ATE} + \frac{\text{NIE}-\text{NIE}^R}{\E(M_{A=1}-M_{A=0})}.
\end{align*}
The first equality is by the independence assumption, the second is by the exclusion assumption (equation 9 in \citet{angrist_identification_1996}), the third is by the definition of covariance, the fourth is by Proposition \ref{prop:binary} and the exclusion assumption. The relevance assumption ensures that the denominator is nonzero. Finally, under assumptions of exclusion, independence, relevance, and monotonicity, the classic result of \citet{angrist_identification_1996} equates Wald with LATE.

\subsection*{A6. Proof of Proposition 7}
\begin{align*}
    &\phantom{{}={}} \E[\one(Y_1 \geq Y_0)] - \E[\one(H_1 \geq H_0)] \\
    &= \iint \one(t \geq s)f_{Y_1,Y_0}(t,s) \dd t \dd s - \iint \one(t \geq s)f_{H_1,H_0}(t,s) \dd t \dd s \\
    &= \iint \one(t \geq s)f_{Y_1,Y_0}(t,s) \dd t \dd s - \iint \one(t \geq s)f_{H_1}(t)f_{H_0}(s) \dd t \dd s \\
    &= \iint \one(t \geq s)f_{Y_1,Y_0}(t,s) \dd t \dd s - \iint \one(t \geq s)f_{Y_1}(t)f_{Y_0}(s) \dd t \dd s \\
    &= \iint \one(t \geq s)\E[\one(Y_1=t)\one(Y_0=s)] \dd t \dd s - \iint \one(t \geq s)\E[\one(Y_1=t)]\E[\one(Y_0=s)] \dd t \dd s \\
    &= \iint \one(t \geq s) \Cov[\one(Y_1 = t),\one(Y_0=s)] \dd t \dd s.
\end{align*}
When $Y$ is discrete, this becomes the expression in Proposition 7. Furthermore, when the support of $Y$ is $\{0,1\}$,
\begin{align*}
    &\phantom{{}={}} \sum_{t \in \mathcal{T}}\sum_{s \in \mathcal{S}} \one(t \geq s) \Cov[\one(Y_1 = t),\one(Y_0=s)] \\
    &= \Cov[\one(Y_1=1),\one(Y_0=1)] + \Cov[\one(Y_1=1),\one(Y_0=0)] + \Cov[\one(Y_1=0),\one(Y_0=0)] \\
    &= \E[\one(Y_1=1)\one(Y_0=1)] - \E[\one(Y_1=1)]\E[\one(Y_0=1)] \\
    &\phantom{{}={}}+ \E[\one(Y_1=1)\one(Y_0=0)] - \E[\one(Y_1=1)]\E[\one(Y_0=0)] \\
    &\phantom{{}={}}+ \E[\one(Y_1=0)\one(Y_0=0)] - \E[\one(Y_1=0)]\E[\one(Y_0=0)] \\
    &= \E(Y_1 Y_0) - \E(Y_1)\E(Y_0) \\
    &\phantom{{}={}}+ \E[Y_1 (1-Y_0)] - \E(Y_1)[1-\E(Y_0)] + \E[(1-Y_1) (1-Y_0)]-\E[(1-Y_1)]\E[(1-Y_0)] \\
    &= \E(Y_1 Y_0) - \E(Y_1)\E(Y_0) = \Cov(Y_1, Y_0).
\end{align*}

\bibliographystyle{chicago}
\bibliography{bibliography.bib}

\begin{thebibliography}{}

\bibitem[\protect\citeauthoryear{Andrews and Didelez}{Andrews and
  Didelez}{2021}]{andrews_insights_2021}
Andrews, R.~M. and V.~Didelez (2021, March).
\newblock Insights into the {Cross}-world {Independence} {Assumption} of
  {Causal} {Mediation} {Analysis}.
\newblock {\em Epidemiology\/}~{\em 32\/}(2), 209--219.

\bibitem[\protect\citeauthoryear{Angrist, Imbens, and Rubin}{Angrist
  et~al.}{1996}]{angrist_identification_1996}
Angrist, J.~D., G.~W. Imbens, and D.~B. Rubin (1996).
\newblock Identification of {Causal} {Effects} {Using} {Instrumental}
  {Variables}.
\newblock {\em Journal of the American Statistical Association\/}~{\em
  91\/}(434), 444--455.

\bibitem[\protect\citeauthoryear{Aronow and Carnegie}{Aronow and
  Carnegie}{2013}]{aronow_beyond_2013}
Aronow, P.~M. and A.~Carnegie (2013).
\newblock Beyond {LATE}: {Estimation} of the {Average} {Treatment} {Effect}
  with an {Instrumental} {Variable}.
\newblock {\em Political Analysis\/}~{\em 21\/}(4), 492--506.

\bibitem[\protect\citeauthoryear{Avin, Shpitser, and Pearl}{Avin
  et~al.}{2005}]{avin_identiability_2005}
Avin, C., I.~Shpitser, and J.~Pearl (2005).
\newblock Identiﬁability of {Path}-{Speciﬁc} {Effects}.
\newblock In {\em Proceedings of {International} {Joint} {Conference} on
  {Artificial} {Intelligence}}, Edinburgh, Schotland, pp.\  357--363.

\bibitem[\protect\citeauthoryear{Balke and Pearl}{Balke and
  Pearl}{1997}]{balke_bounds_1997}
Balke, A. and J.~Pearl (1997, September).
\newblock Bounds on {Treatment} {Effects} {From} {Studies} {With} {Imperfect}
  {Compliance}.
\newblock {\em Journal of the American Statistical Association\/}~{\em
  92\/}(439), 1171--1176.

\bibitem[\protect\citeauthoryear{Baron and Kenny}{Baron and
  Kenny}{1986}]{baron_moderatormediator_1986}
Baron, R.~M. and D.~A. Kenny (1986).
\newblock The moderator–mediator variable distinction in social psychological
  research: {Conceptual}, strategic, and statistical considerations.
\newblock {\em Journal of personality and social psychology\/}~{\em 51\/}(6),
  1173--1182.

\bibitem[\protect\citeauthoryear{Chernozhukov, Newey, Quintas-Martinez, and
  Syrgkanis}{Chernozhukov et~al.}{2024}]{chernozhukov_2024}
Chernozhukov, V., W.~K. Newey, V.~Quintas-Martinez, and V.~Syrgkanis (2024).
\newblock Automatic debiased machine learning via riesz regression.

\bibitem[\protect\citeauthoryear{Demidenko}{Demidenko}{2016}]{demidenko_p_2016}
Demidenko, E. (2016, January).
\newblock The \textit{p} -{Value} {You} {Can}’t {Buy}.
\newblock {\em The American Statistician\/}~{\em 70\/}(1), 33--38.

\bibitem[\protect\citeauthoryear{Díaz, Hejazi, Rudolph, and Van
  Der~Laan}{Díaz et~al.}{2021}]{diaz_nonparametric_2021}
Díaz, I., N.~S. Hejazi, K.~E. Rudolph, and M.~J. Van Der~Laan (2021, August).
\newblock Nonparametric efficient causal mediation with intermediate
  confounders.
\newblock {\em Biometrika\/}~{\em 108\/}(3), 627--641.

\bibitem[\protect\citeauthoryear{Fay, Brittain, Shih, Follmann, and
  Gabriel}{Fay et~al.}{2018}]{fay_causal_2018}
Fay, M.~P., E.~H. Brittain, J.~H. Shih, D.~A. Follmann, and E.~E. Gabriel
  (2018, September).
\newblock Causal estimands and confidence intervals associated with
  {Wilcoxon}‐{Mann}‐{Whitney} tests in randomized experiments.
\newblock {\em Statistics in Medicine\/}~{\em 37\/}(20), 2923--2937.

\bibitem[\protect\citeauthoryear{Glynn}{Glynn}{2012}]{glynn_product_2012}
Glynn, A.~N. (2012, January).
\newblock The {Product} and {Difference} {Fallacies} for {Indirect} {Effects}.
\newblock {\em American Journal of Political Science\/}~{\em 56\/}(1),
  257--269.

\bibitem[\protect\citeauthoryear{Greenland, Fay, Brittain, Shih, Follmann,
  Gabriel, and Robins}{Greenland et~al.}{2020}]{greenland_causal_2020}
Greenland, S., M.~P. Fay, E.~H. Brittain, J.~H. Shih, D.~A. Follmann, E.~E.
  Gabriel, and J.~M. Robins (2020, July).
\newblock On {Causal} {Inferences} for {Personalized} {Medicine}: {How}
  {Hidden} {Causal} {Assumptions} {Led} to {Erroneous} {Causal} {Claims}
  {About} the \textit{{D}} -{Value}.
\newblock {\em The American Statistician\/}~{\em 74\/}(3), 243--248.

\bibitem[\protect\citeauthoryear{Hand}{Hand}{1992}]{hand_comparing_1992}
Hand, D.~J. (1992, August).
\newblock On {Comparing} {Two} {Treatments}.
\newblock {\em The American Statistician\/}~{\em 46\/}(3), 190--192.

\bibitem[\protect\citeauthoryear{Hartwig, Wang, Smith, and Davies}{Hartwig
  et~al.}{2023}]{hartwig_average_2023}
Hartwig, F.~P., L.~Wang, G.~D. Smith, and N.~M. Davies (2023).
\newblock Average causal effect estimation via instrumental variables: the no
  simultaneous heterogeneity assumption.
\newblock {\em Epidemiology\/}~{\em 34\/}(3), 325--332.

\bibitem[\protect\citeauthoryear{Hayden, Pauler, and Schoenfeld}{Hayden
  et~al.}{2005}]{hayden_estimator_2005}
Hayden, D., D.~K. Pauler, and D.~Schoenfeld (2005, March).
\newblock An {Estimator} for {Treatment} {Comparisons} among {Survivors} in
  {Randomized} {Trials}.
\newblock {\em Biometrics\/}~{\em 61\/}(1), 305--310.

\bibitem[\protect\citeauthoryear{Heckman}{Heckman}{1997}]{heckman_instrumental_1997}
Heckman, J. (1997).
\newblock Instrumental {Variables}: {A} {Study} of {Implicit} {Behavioral}
  {Assumptions} {Used} in {Making} {Program} {Evaluations}.
\newblock {\em The Journal of Human Resources\/}~{\em 32\/}(3), 441.

\bibitem[\protect\citeauthoryear{Hernán and Robins}{Hernán and
  Robins}{2006}]{hernan_instruments_2006}
Hernán, M.~A. and J.~M. Robins (2006, July).
\newblock Instruments for {Causal} {Inference}: {An} {Epidemiologist}'s
  {Dream}?
\newblock {\em Epidemiology\/}~{\em 17\/}(4), 360--372.

\bibitem[\protect\citeauthoryear{Hernán and Robins}{Hernán and
  Robins}{2020}]{hernan_causal_2020}
Hernán, M.~A. and J.~M. Robins (2020).
\newblock {\em Causal {Inference}: {What} {If}}.
\newblock Boca Raton: Chapman \& Hall/CRC.

\bibitem[\protect\citeauthoryear{{Holland}}{{Holland}}{1986}]{holland_1986}
{Holland}, P. (1986, 12).
\newblock Statistics and causal inference.
\newblock {\em Journal of the American Statistical Association\/}~{\em
  81\/}(396), 945--960.

\bibitem[\protect\citeauthoryear{Hong}{Hong}{2015}]{hong2015causality}
Hong, G. (2015).
\newblock {\em Causality in a social world: Moderation, mediation and
  spill-over}.
\newblock John Wiley \& Sons.

\bibitem[\protect\citeauthoryear{Imai, Keele, and Yamamoto}{Imai
  et~al.}{2010}]{imai_identification_2010}
Imai, K., L.~Keele, and T.~Yamamoto (2010, February).
\newblock Identification, {Inference} and {Sensitivity} {Analysis} for {Causal}
  {Mediation} {Effects}.
\newblock {\em Statistical Science\/}~{\em 25\/}(1), 51 -- 71.

\bibitem[\protect\citeauthoryear{Imbens}{Imbens}{2010}]{Imbens_2010}
Imbens, G.~W. (2010, June).
\newblock Better late than nothing: Some comments on deaton (2009) and heckman
  and urzua (2009).
\newblock {\em Journal of Economic Literature\/}~{\em 48\/}(2), 399–423.

\bibitem[\protect\citeauthoryear{Imbens and Rubin}{Imbens and
  Rubin}{2015}]{imbens2015causal}
Imbens, G.~W. and D.~B. Rubin (2015).
\newblock {\em Causal inference in statistics, social, and biomedical
  sciences}.
\newblock Cambridge university press.

\bibitem[\protect\citeauthoryear{Kling, Ludwig, and Katz}{Kling
  et~al.}{2005}]{kling_neighborhood_2005}
Kling, J.~R., J.~Ludwig, and L.~F. Katz (2005, February).
\newblock Neighborhood {Effects} on {Crime} for {Female} and {Male} {Youth}:
  {Evidence} from a {Randomized} {Housing} {Voucher} {Experiment}.
\newblock {\em The Quarterly Journal of Economics\/}~{\em 120\/}(1), 87--130.

\bibitem[\protect\citeauthoryear{Lei}{Lei}{2024}]{lei_causal_2024}
Lei, L. (2024, June).
\newblock Causal {Interpretation} of {Regressions} {With} {Ranks}.
\newblock arXiv:2406.05548 [econ, math, stat].

\bibitem[\protect\citeauthoryear{Liu, Williams, Rudolph, and Díaz}{Liu
  et~al.}{2024}]{liu_general_2024}
Liu, R., N.~T. Williams, K.~E. Rudolph, and I.~Díaz (2024, August).
\newblock General targeted machine learning for modern causal mediation
  analysis.
\newblock arXiv:2408.14620 [cs, stat].

\bibitem[\protect\citeauthoryear{Loh, Moerkerke, Loeys, and Vansteelandt}{Loh
  et~al.}{2020}]{Loh_2020}
Loh, W.~W., B.~Moerkerke, T.~Loeys, and S.~Vansteelandt (2020).
\newblock Heterogeneous indirect effects for multiple mediators using
  interventional effect models.
\newblock {\em Epidemiologic Methods\/}~{\em 9\/}(1), 20200023.

\bibitem[\protect\citeauthoryear{Lok}{Lok}{2016}]{lok_defining_2016}
Lok, J.~J. (2016, September).
\newblock Defining and estimating causal direct and indirect effects when
  setting the mediator to specific values is not feasible.
\newblock {\em Statistics in Medicine\/}~{\em 35\/}(22), 4008--4020.

\bibitem[\protect\citeauthoryear{Lok and Bosch}{Lok and
  Bosch}{2021}]{lok_causal_2021}
Lok, J.~J. and R.~J. Bosch (2021, May).
\newblock Causal {Organic} {Indirect} and {Direct} {Effects}: {Closer} to the
  {Original} {Approach} to {Mediation} {Analysis}, with a {Product} {Method}
  for {Binary} {Mediators}.
\newblock {\em Epidemiology\/}~{\em 32\/}(3), 412--420.

\bibitem[\protect\citeauthoryear{Lu, Zhang, and Ding}{Lu et~al.}{2020}]{lu2020}
Lu, J., Y.~Zhang, and P.~Ding (2020).
\newblock Sharp bounds on the relative treatment effect for ordinal outcomes.
\newblock {\em Biometrics\/}~{\em 76\/}(2), 664--669.

\bibitem[\protect\citeauthoryear{Ludwig, Duncan, Gennetian, Katz, Kessler,
  Kling, and Sanbonmatsu}{Ludwig et~al.}{2013}]{ludwig_neighborhood_2013}
Ludwig, J., G.~J. Duncan, L.~A. Gennetian, L.~R. Katz, R.~Kessler, J.~R. Kling,
  and L.~Sanbonmatsu (2013, March).
\newblock Neighborhood {Effects} on the {Long}-{Term} {Well}-{Being} of
  {Low}-{Income} {Adults} {From} {All} {Five} {Sites} of the {Moving} to
  {Opportunity} {Experiment}, 2008-2010 [{Public} {Use} {Data}].

\bibitem[\protect\citeauthoryear{Mann and Whitney}{Mann and
  Whitney}{1947}]{mann_whitney_test_1947}
Mann, H.~B. and D.~R. Whitney (1947).
\newblock On a test of whether one of two random variables is stochastically
  larger than the other.
\newblock {\em The annals of mathematical statistics\/}~{\em 18\/}(1), 50--60.

\bibitem[\protect\citeauthoryear{Mao}{Mao}{2018}]{mao_causal_2018}
Mao, L. (2018, March).
\newblock On causal estimation using u-statistics.
\newblock {\em Biometrika\/}~{\em 105\/}(1), 215--220.

\bibitem[\protect\citeauthoryear{Mao}{Mao}{2024}]{mao_wilcoxon-mann-whitney_2024}
Mao, L. (2024, January).
\newblock Wilcoxon-{Mann}-{Whitney} statistics in randomized trials with
  non-compliance.
\newblock {\em Electronic Journal of Statistics\/}~{\em 18\/}(1).

\bibitem[\protect\citeauthoryear{Miles, Kanki, Meloni, and
  Tchetgen~Tchetgen}{Miles et~al.}{2017}]{miles_partial_2017}
Miles, C., P.~Kanki, S.~Meloni, and E.~Tchetgen~Tchetgen (2017, February).
\newblock On {Partial} {Identification} of the {Natural} {Indirect} {Effect}.
\newblock {\em Journal of Causal Inference\/}~{\em 5\/}(2), 20160004.

\bibitem[\protect\citeauthoryear{Miles}{Miles}{2023}]{miles_causal_2023}
Miles, C.~H. (2023, September).
\newblock On the causal interpretation of randomised interventional indirect
  effects.
\newblock {\em Journal of the Royal Statistical Society Series B: Statistical
  Methodology\/}~{\em 85\/}(4), 1154--1172.

\bibitem[\protect\citeauthoryear{Nguyen, Schmid, Ogburn, and Stuart}{Nguyen
  et~al.}{2022}]{nguyen_clarifying_2022}
Nguyen, T.~Q., I.~Schmid, E.~L. Ogburn, and E.~A. Stuart (2022, September).
\newblock Clarifying causal mediation analysis: {Effect} identification via
  three assumptions and five potential outcomes.
\newblock {\em Journal of Causal Inference\/}~{\em 10\/}(1), 246--279.

\bibitem[\protect\citeauthoryear{Nguyen, Schmid, and Stuart}{Nguyen
  et~al.}{2021}]{nguyen_clarifying_2021}
Nguyen, T.~Q., I.~Schmid, and E.~A. Stuart (2021, April).
\newblock Clarifying causal mediation analysis for the applied researcher:
  {Defining} effects based on what we want to learn.
\newblock {\em Psychological Methods\/}~{\em 26\/}(2), 255--271.

\bibitem[\protect\citeauthoryear{Pearl}{Pearl}{1995}]{pearl_causal_1995}
Pearl, J. (1995, December).
\newblock Causal {Diagrams} for {Empirical} {Research}.
\newblock {\em Biometrika\/}~{\em 82\/}(4), 669--688.

\bibitem[\protect\citeauthoryear{Pearl}{Pearl}{2001}]{pearl_direct_2001}
Pearl, J. (2001).
\newblock Direct and {Indirect} {Effects}.
\newblock In {\em Proceedings of the {Seventeenth} {Conference} on {Uncertainy}
  in {Artificial} {Intel} ligence}, San Francisco, CA, pp.\  411--20. Morgan
  Kaufmann.

\bibitem[\protect\citeauthoryear{Pearl}{Pearl}{2012}]{pearl_causal_2012}
Pearl, J. (2012).
\newblock The {Causal} {Foundations} of {Structural} {Equation} {Modeling}:.
\newblock In R.~H. Hoyle (Ed.), {\em Handbook of {Structural} {Equation}
  {Modeling}}, pp.\  68--91. The Guilford Press.

\bibitem[\protect\citeauthoryear{Pocock, Ariti, Collier, and Wang}{Pocock
  et~al.}{2012}]{pocock2012}
Pocock, S.~J., C.~A. Ariti, T.~J. Collier, and D.~Wang (2012).
\newblock The win ratio: a new approach to the analysis of composite endpoints
  in clinical trials based on clinical priorities.
\newblock {\em European heart journal\/}~{\em 33\/}(2), 176--182.

\bibitem[\protect\citeauthoryear{Qu, Fu, Luo, and Ruberg}{Qu
  et~al.}{2020}]{qu_general_2020}
Qu, Y., H.~Fu, J.~Luo, and S.~J. Ruberg (2020, January).
\newblock A {General} {Framework} for {Treatment} {Effect} {Estimators}
  {Considering} {Patient} {Adherence}.
\newblock {\em Statistics in Biopharmaceutical Research\/}~{\em 12\/}(1),
  1--18.

\bibitem[\protect\citeauthoryear{Robins}{Robins}{2003}]{green_semantics_2003}
Robins, J.~M. (2003, May).
\newblock Semantics of causal {DAG} models and the identification of direct and
  indirect effects.
\newblock In P.~J. Green, N.~L. Hjort, and S.~Richardson (Eds.), {\em Highly
  {Structured} {Stochastic} {Systems}}, pp.\  70--82. Oxford: Oxford University
  Press.

\bibitem[\protect\citeauthoryear{Robins and Greenland}{Robins and
  Greenland}{1992}]{robins1992}
Robins, J.~M. and S.~Greenland (1992).
\newblock Identifiability and exchangeability for direct and indirect effects.
\newblock {\em Epidemiology\/}~{\em 3\/}(2), 143--155.

\bibitem[\protect\citeauthoryear{Robins and Greenland}{Robins and
  Greenland}{1996}]{robins_identification_1996}
Robins, J.~M. and S.~Greenland (1996).
\newblock Identification of causal effects using instrumental variables:
  comment.
\newblock {\em Journal of the American Statistical Association\/}~{\em
  91\/}(434), 456--458.

\bibitem[\protect\citeauthoryear{Rudolph, Gimbrone, and Díaz}{Rudolph
  et~al.}{2021}]{rudolph_helped_2021}
Rudolph, K.~E., C.~Gimbrone, and I.~Díaz (2021, May).
\newblock Helped into {Harm}: {Mediation} of a {Housing} {Voucher}
  {Intervention} on {Mental} {Health} and {Substance} {Use} in {Boys}.
\newblock {\em Epidemiology\/}~{\em 32\/}(3), 336--346.

\bibitem[\protect\citeauthoryear{Rudolph, Williams, and Diaz}{Rudolph
  et~al.}{2024}]{rudolph_practical_2024}
Rudolph, K.~E., N.~T. Williams, and I.~Diaz (2024, April).
\newblock Practical causal mediation analysis: extending nonparametric
  estimators to accommodate multiple mediators and multiple intermediate
  confounders.
\newblock {\em Biostatistics\/}, kxae012.

\bibitem[\protect\citeauthoryear{Sarvet, Stensrud, and Wen}{Sarvet
  et~al.}{2023}]{sarvet_interpretational_2023}
Sarvet, A.~L., M.~J. Stensrud, and L.~Wen (2023, December).
\newblock Interpretational errors in statistical causal inference.
\newblock arXiv:2312.07610 [stat].

\bibitem[\protect\citeauthoryear{Swanson, Hernán, Miller, Robins, and
  Richardson}{Swanson et~al.}{2018}]{swanson_partial_2018}
Swanson, S.~A., M.~A. Hernán, M.~Miller, J.~M. Robins, and T.~S. Richardson
  (2018, April).
\newblock Partial {Identification} of the {Average} {Treatment} {Effect}
  {Using} {Instrumental} {Variables}: {Review} of {Methods} for {Binary}
  {Instruments}, {Treatments}, and {Outcomes}.
\newblock {\em Journal of the American Statistical Association\/}~{\em
  113\/}(522), 933--947.

\bibitem[\protect\citeauthoryear{Thas}{Thas}{2010}]{thas_comparing_2010}
Thas, O. (2010).
\newblock {\em Comparing distributions}.
\newblock Springer {Series} in {Statistics}. New York: Springer.

\bibitem[\protect\citeauthoryear{Thas, Neve, Clement, and Ottoy}{Thas
  et~al.}{2012}]{thas_probabilistic_2012}
Thas, O., J.~D. Neve, L.~Clement, and J.-P. Ottoy (2012, September).
\newblock Probabilistic {Index} {Models}.
\newblock {\em Journal of the Royal Statistical Society Series B: Statistical
  Methodology\/}~{\em 74\/}(4), 623--671.

\bibitem[\protect\citeauthoryear{Tian and Pearl}{Tian and
  Pearl}{2000}]{tian_probabilities_2000}
Tian, J. and J.~Pearl (2000).
\newblock Probabilities of causation: {Bounds} and identification.
\newblock {\em Annals of Mathematics and Artificial Intelligence\/}~{\em
  28\/}(1), 287--313.

\bibitem[\protect\citeauthoryear{VanderWeele}{VanderWeele}{2015}]{vanderweele_explanation_2015}
VanderWeele, T. (2015).
\newblock {\em Explanation in Causal Inference: Methods for Mediation and
  Interaction}.
\newblock Oxford University Press.

\bibitem[\protect\citeauthoryear{VanderWeele and Tchetgen~Tchetgen}{VanderWeele
  and Tchetgen~Tchetgen}{2017}]{vanderweele_mediation_2017}
VanderWeele, T.~J. and E.~J. Tchetgen~Tchetgen (2017, June).
\newblock Mediation analysis with time varying exposures and mediators.
\newblock {\em Journal of the Royal Statistical Society: Series B (Statistical
  Methodology)\/}~{\em 79\/}(3), 917--938.

\bibitem[\protect\citeauthoryear{VanderWeele, Vansteelandt, and
  Robins}{VanderWeele et~al.}{2014}]{vanderweele_effect_2014}
VanderWeele, T.~J., S.~Vansteelandt, and J.~M. Robins (2014, March).
\newblock Effect {Decomposition} in the {Presence} of an {Exposure}-{Induced}
  {Mediator}-{Outcome} {Confounder}.
\newblock {\em Epidemiology\/}~{\em 25\/}(2), 300--306.

\bibitem[\protect\citeauthoryear{Vansteelandt and Daniel}{Vansteelandt and
  Daniel}{2017}]{vansteelandt_interventional_2017}
Vansteelandt, S. and R.~M. Daniel (2017, March).
\newblock Interventional {Effects} for {Mediation} {Analysis} with {Multiple}
  {Mediators}:.
\newblock {\em Epidemiology\/}~{\em 28\/}(2), 258--265.

\bibitem[\protect\citeauthoryear{Vansteelandt and Lancker}{Vansteelandt and
  Lancker}{2024}]{vansteelandt_chasing_2024}
Vansteelandt, S. and K.~V. Lancker (2024, October).
\newblock Chasing {Shadows}: {How} {Implausible} {Assumptions} {Skew} {Our}
  {Understanding} of {Causal} {Estimands}.
\newblock arXiv:2409.11162 [stat].

\bibitem[\protect\citeauthoryear{Wang and Tchetgen~Tchetgen}{Wang and
  Tchetgen~Tchetgen}{2018}]{wang_bounded_2018}
Wang, L. and E.~Tchetgen~Tchetgen (2018, June).
\newblock Bounded, {Efficient} and {Multiply} {Robust} {Estimation} of
  {Average} {Treatment} {Effects} {Using} {Instrumental} {Variables}.
\newblock {\em Journal of the Royal Statistical Society Series B: Statistical
  Methodology\/}~{\em 80\/}(3), 531--550.

\bibitem[\protect\citeauthoryear{Wilcoxon}{Wilcoxon}{1945}]{wilcoxon_individual_1945}
Wilcoxon, F. (1945).
\newblock Individual {Comparisons} by {Ranking} {Methods}.
\newblock {\em Biometrics Bulletin\/}~{\em 1\/}(6), 80--83.

\bibitem[\protect\citeauthoryear{Williams and Díaz}{Williams and
  Díaz}{2024}]{Williams_2024}
Williams, N. and I.~Díaz (2024).
\newblock {\em crumble: Flexible and General Mediation Analysis Using Riesz
  Representers}.
\newblock R package version 0.1.0, commit
  9240ff8c9ff2c31f257abb9f2bc88eba843bae9e.

\bibitem[\protect\citeauthoryear{Wodtke and Zhou}{Wodtke and
  Zhou}{2020}]{wodtke_effect_2020}
Wodtke, G.~T. and X.~Zhou (2020, May).
\newblock Effect {Decomposition} in the {Presence} of {Treatment}-induced
  {Confounding}: {A} {Regression}-with-residuals {Approach}.
\newblock {\em Epidemiology\/}~{\em 31\/}(3), 369--375.

\bibitem[\protect\citeauthoryear{Wu, Han, Chen, and Tu}{Wu
  et~al.}{2014}]{wu_causal_2014}
Wu, P., Y.~Han, T.~Chen, and X.~Tu (2014, April).
\newblock Causal inference for {Mann}-{Whitney}-{Wilcoxon} rank sum and other
  nonparametric statistics.
\newblock {\em Statistics in Medicine\/}~{\em 33\/}(8), 1261--1271.

\bibitem[\protect\citeauthoryear{Yu and Elwert}{Yu and
  Elwert}{2025}]{yu2025nonparametric}
Yu, A. and F.~Elwert (2025).
\newblock Nonparametric causal decomposition of group disparities.
\newblock {\em The Annals of Applied Statistics\/}~{\em 19\/}(1), 821--845.

\bibitem[\protect\citeauthoryear{Zheng and Van Der~Laan}{Zheng and Van
  Der~Laan}{2017}]{zheng_longitudinal_2017}
Zheng, W. and M.~Van Der~Laan (2017, June).
\newblock Longitudinal {Mediation} {Analysis} with {Time}-varying {Mediators}
  and {Exposures}, with {Application} to {Survival} {Outcomes}.
\newblock {\em Journal of Causal Inference\/}~{\em 5\/}(2), 20160006.

\bibitem[\protect\citeauthoryear{Zhou}{Zhou}{2022}]{zhou_attendance_2022}
Zhou, X. (2022, August).
\newblock Attendance, {Completion}, and {Heterogeneous} {Returns} to {College}:
  {A} {Causal} {Mediation} {Approach}.
\newblock {\em Sociological Methods \& Research\/}, 004912412211138.

\end{thebibliography}

\end{document}